\documentclass[acmsmall]{acmart}

\usepackage{makecell}
\usepackage{hhline}
\usepackage{mathtools, nccmath}
\usepackage{booktabs}
\usepackage{verbatimbox}
\usepackage{amsmath,amssymb,amsfonts}
\DeclareMathOperator*{\argmax}{arg\,max}
\DeclareMathOperator*{\argmin}{arg\,min}

\usepackage{threeparttable} 
\usepackage{algorithmic}
\usepackage[ruled,vlined,linesnumbered]{algorithm2e}

\SetCommentSty{mycommfont}
\usepackage{float}
\usepackage{graphicx}
\usepackage{hyperref}  
\hypersetup{
    colorlinks=true,    
    linkcolor=black,    
    citecolor=black,    
    urlcolor=black,     
    filecolor=black,    
}
\usepackage{multirow}
\usepackage{subfiles}
\usepackage{subfigure}
\usepackage{subcaption}
\usepackage{textcomp}
\usepackage{amsthm}
\usepackage{bm}
\usepackage{url}

\newtheorem{definition}{Definition}
\newcommand{\add}[1]{{\leavevmode\color{blue}#1}}

\AtBeginDocument{%
  }

\acmJournal{TOSN}

\begin{document}
\title{VTarbel: Targeted Label Attack with Minimal Knowledge
  on Detector-enhanced Vertical Federated Learning
}

\author{Juntao Tan}
\affiliation{%
  \institution{University of Science and Technology of China}
  \city{Hefei}
  \country{China}
}
\email{tjt@mail.ustc.edu.cn}

\author{Anran Li}
\affiliation{
  \institution{University of Science and Technology of China}
  \city{Hefei}
  \country{China}
}
\email{anranli@mail.ustc.edu.cn}

\author{Quanchao Liu}
\affiliation{
  \institution{Department of Security Technology Research, China Mobile Research Institute}
  \city{Beijing}
  \country{China}
}
\email{liuquanchao@chinamobile.com}

\author{Peng Ran}
\affiliation{
  \institution{Department of Security Technology Research, China Mobile Research Institute}
  \city{Beijing}
  \country{China}
}
\email{ranpeng@chinamobile.com}

\author{Lan Zhang}
\affiliation{
  \institution{University of Science and Technology of China}
  \city{Hefei}
  \country{China}
}
\email{zhanglan@ustc.edu.cn}


\renewcommand{\shortauthors}{Tan et al.}

\begin{abstract}
Vertical federated learning (VFL) enables multiple parties with disjoint features to collaboratively train models without sharing raw data. 
While privacy vulnerabilities of VFL are extensively-studied, its security threats—particularly targeted label attacks—remain underexplored. 
In such attacks, a passive party perturbs inputs at inference to force misclassification into adversary-chosen labels. 
Existing methods rely on unrealistic assumptions (\emph{e.g.}, accessing VFL-model's outputs) and ignore anomaly detectors deployed in real-world systems.
To bridge this gap, we introduce VTarbel, a two-stage, minimal-knowledge attack framework explicitly designed to evade detector-enhanced VFL inference.
During the preparation stage, the attacker selects a minimal set of high-expressiveness samples (via maximum mean discrepancy), submits them through VFL protocol to collect predicted labels, and uses these pseudo-labels to train estimated detector and surrogate model on local features. 
In attack stage, these models guide gradient-based perturbations of remaining samples, crafting adversarial instances that induce targeted misclassifications and evade detection.
We implement VTarbel and evaluate it against four model architectures, seven multimodal datasets, and two anomaly detectors. 
Across all settings, VTarbel outperforms four state-of-the-art baselines, evades detection, and retains effective against three representative privacy-preserving defenses. 
These results reveal critical security blind spots in current VFL deployments and underscore urgent need for robust, attack-aware defenses.

\end{abstract}

\ccsdesc[500]{Security and privacy~Distributed systems security}
\ccsdesc[500]{Computing methodologies~Distributed artificial intelligence}

\keywords{Vertical Federated Learning, Targeted Label Attack, Adversarial Attack, Security Risk}


\maketitle

\section{Introduction}


Vertical Federated Learning (VFL) is a decentralized machine learning (ML) paradigm that enables collaborative model training among multiple participants, each holding distinct, non-overlapping feature sets, without sharing their raw data ~\cite{yang2019federated,liu2022vertical,li2023vertical,yang2023survey,wei2022vertical,khan2022vertical,jia2024model}. 
Typically, one participant, referred to as the \emph{active party}, possesses the ground-truth labels, while the others, referred to as \emph{passive parties}, contribute only raw features. 
VFL has gained traction in real-world applications such as financial risk assessment ~\cite{chen2021homomorphic}, healthcare diagnostics ~\cite{yang2019federated}, and personalized advertising ~\cite{li2022vertical}.

To protect data privacy, VFL protocols restrict information exchange to intermediate outputs such as feature embeddings\footnote{Feature embeddings denote the intermediate representations generated by deep learning models.} and embedding gradients. 
Despite these protections, recent studies have shown that VFL is still vulnerable to privacy leakage via two main types of attacks. 
Feature reconstruction attacks allow the active party to recover the passive parties' original features from the shared embeddings ~\cite{he2019model,luo2021feature,weng2020privacy,jin2021cafe,gao2023pcat,pasquini2021unleashing,fu2023focusing}, while label inference attacks enable malicious passive parties to deduce private labels using backward gradients or their locally trained models ~\cite{fu2022label,li2022label,sun2022label,zou2022label,tan2022residue,kang2022framework}. 

Despite growing attention to privacy breaches in VFL, its robustness against adversarial perturbations remain  underexplored. 
A particularly critical but overlooked threat is \emph{targeted label attacks}, in which a passive party (the attacker) deliberately crafts malicious input samples to mislead the VFL model into producing attacker-specified misclassifications. 
For instance, an adversary in a healthcare VFL system could manipulate diagnostic outputs to misguide treatments.
Such attacks generally fall into two categories: 
(1) Training-phase backdoor attacks \cite{naseri2024badvfl,bai2023villain,liu2024pna}, where attackers poison training data to implant hidden triggers, once activated during inference, cause targeted  misclassifications.
(2) Inference-phase adversarial attacks \cite{gu2023lrba,qiu2024hijack,pang2023adi,he2023backdoor,liu2022copur}, where attackers directly perturb test inputs to manipulate model predictions without modifying the training process (as illustrated in Figure \ref{fig:vfl_inference_system}).
While both types pose security risks, backdoor attacks often degrade overall model performance due to the presence of poisoned samples, potentially alerting the active party (the defender) and lead to the early termination of the collaboration. 
In contrast, adversarial attacks operate stealthily without altering training data, which significantly increases the difficulty of detection. 
Given their practicality and stealthiness, this work focuses on inference-phase targeted label attacks (hereafter refer to as targeted label attacks for brevity).

Designing effective targeted label attacks in VFL during inference  is highly challenging due to three key factors. 
First, the attacker typically operates in a black-box setting without access to the defender's model architecture, parameters, or outputs, which severely limits their ability to manipulate predictions. 
Second, the attacker controls only a subset of the features, which means that adversarial perturbations may be mitigated by benign features contributed by other participants. 
Third, adversarial samples may be detected by anomaly detection mechanisms deployed by the defender to monitor input distributions.


Several recent studies have proposed attack strategies to address these challenges. For example, Gu et al. \cite{gu2023lrba} and Qiu et al. \cite{qiu2024hijack} leverage auxiliary labeled datasets to train auxiliary models, which then generate adversarial embeddings by aligning them with target-class representations. 
Pang et al. \cite{pang2023adi} assume the attacker has access to test samples from other participants as well as the VFL model's outputs, allowing iterative optimization of adversarial inputs. 
He et al. \cite{he2023backdoor} propose using labeled target-class samples to extract representative embeddings that are injected into test samples to mislead  predictions.
However, existing targeted label attacks have critical limitations in practical applications. 
\emph{(1) Unrealistic adversarial knowledge:} Many proposed methods rely on strong assumptions, such as access to labeled datasets, global model outputs, or test samples from other participants, which do not hold in real VFL scenarios. In practice, labels are proprietary assets exclusively held by the active party and are never shared with passive parties \cite{chen2021homomorphic}.
\emph{(2) Neglect of anomaly detection:} These approaches often ignore the active party's ability to deploy anomaly detectors, \emph{e.g.}, statistical or ML-based, to identify and filter out suspicious embeddings.
As a result, the adversarial feature embeddings crafted by attackers are often easily detected, severely compromising attack effectiveness.

To address these issues, we propose a practical and stealthy targeted label attack framework that operates under minimal adversarial knowledge and remains effective against VFL systems equipped with anomaly detectors.
Our design is motivated by empirical observations: overly optimized feature embeddings produced by prior methods tend to deviate significantly from the benign distribution, making them highly detectable and thereby diminishing attack success rates (ASR), as demonstrated in our experimental results (see Section \ref{sec:observations}). 
Based on this insight, we introduce a two-stage attack paradigm. 
In the \emph{preparation stage}, the attacker follows the standard VFL inference protocol to gather essential knowledge, such as estimating the defender's anomaly detection boundaries.
In the \emph{attack stage}, the attacker leverages this knowledge to craft adversarial inputs that both induce targeted misclassifications and remain within the normal range of detector-accepted embeddings, effectively evading detection.


We formally formulate the targeted label attack as a combinatorial optimization problem: selecting an optimal subset of test samples that maximizes the number of predictions assigned to a specific targeted label.
However, solving this NP-hard problem is computationally infeasible. 
To address this, we propose \textbf{VTarbel} (\textbf{\underline{V}}FL Inference \textbf{\underline{Tar}}geted La\textbf{\underline{bel}} Attack), a novel and practical two-stage attack framework designed to operate under realistic VFL constraints. 
VTarbel strategically minimizes the use of benign samples in the preparation stage, reserving the majority of test samples for adversarial crafting during the attack stage. 
In preparation, the attacker selects a small but representative samples set of benign samples based on maximum mean discrepancy (MMD), ensuring high expressiveness and diversity. 
These samples are submitted to the VFL system following standard inference protocol. The corresponding predicted labels are used as pseudo-labels to locally  train two critical components: 
(1) an estimated anomaly detector that approximate the defender's detection logic, and (2) a surrogate model that mimics the behavior of the global VFL model. 
In the attack stage, the attacker applies gradient-based optimization to perturb the remaining test samples. These perturbations are guided by predictions from the surrogate model and constrained by the estimated detector, ensuring that the crafted adversarial samples both induce targeted misclassifications and evade detection.
By leveraging the transferability of adversarial examples, the attacker can effectively bypass the defender's detector and manipulate predictions within the VFL system. 

Our main contributions are summarized as follows.

\begin{itemize}
    \item {\bf Formalizing targeted label attack as a combinatorial optimization problem.} We formalize the targeted label attack in VFL as a combinatorial optimization task, aiming to maximize the number of test samples misclassified into an attacker-specified label. To enable this, we propose a novel inference-phase partitioning that separates the attack process into two sub-stages. 

    \item {\bf Two-stage attack framework with minimal adversarial knowledge.} We introduce {\bf VTarbel}, a detector-agnostic two-stage attack paradigm: (1) a preparation stage that minimizes the required samples to train expressive estimated detector and surrogate model, and (2) an attack stage that generates transferable malicious samples through coordinated optimization of these two models, enabling effective attacks against VFL systems. 

    \item {\bf Extensive evaluations and robustness to defenses.} We conduct extensive evaluations across four model architectures (MLP3, VGG13, ResNet18, DistilBERT), seven multimodal datasets, and two anomaly detectors. Results  demonstrate that VTarbel consistently outperforms four state-of-the-art attacks (\emph{e.g.}, achieving 90.39\% ASR on VGG13 vs. $ \leq$ 49.62\% for baselines). 
    Furthermore, we demonstrate VTarbel's robustness against various defense mechanisms—ASR is reduced by <81.8\% only at the cost of >82.3\% inference accuracy degradation. 
    
\end{itemize}

\label{sec:intro}

\section{Background}

\subsection{Detector-enhanced VFL Inference System} \label{sec:vfl_inference_system}

\begin{figure}
    \centering
    \includegraphics[width=0.85\linewidth]{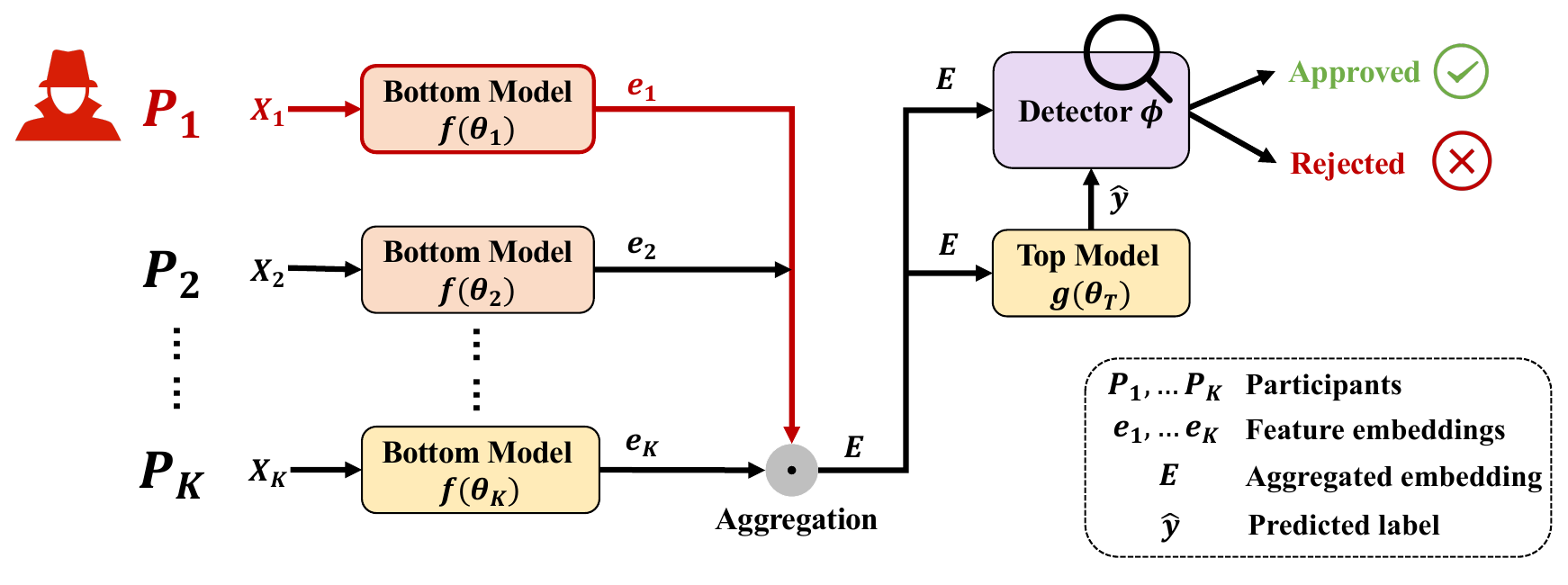}
    \caption{
        Illustration of a detector-enhanced VFL inference system. 
        A malicious passive party (\textcolor{red}{$P_1$}) submits an adversarial feature embedding (\textcolor{red}{$e_1$}) to the active party $P_K$. 
        The active party employs an anomaly detector $\phi$ to assess whether the aggregated embedding $E$ is anomalous. 
        Based on the detector's output, the final prediction is either \textcolor{green}{approved} (\(\hat{y}\)) or \textcolor{red}{rejected} (\(\texttt{REJ}\)).  
    }
    \label{fig:vfl_inference_system}
\end{figure}

A detector-enhanced VFL inference system, as depicted in Figure \ref{fig:vfl_inference_system}, comprises \( K \) participants \( P_1, P_2, \ldots, P_K \). 
We assume \( P_K \) serves as the only active party, while the remaining \( K-1 \) participants act as passive parties. 
Each participant \( P_k \) (\( k \in [K] \), where \( [K] = \{1, 2, \ldots, K\} \)) possesses a private test dataset \( D_k = \{X_k^i\}_{i=1}^N \), where \( X_k^i \) denotes the feature vector of the \( i \)-th test sample for \( P_k \), and \( N \) represents the total number of test samples.  

Following VFL training, the VFL model \( F(\boldsymbol{\theta}) \) is parameterized as \( F(\theta_1, \theta_2, \ldots, \theta_K, \theta_T) \). Here, \( \theta_1, \theta_2, \ldots, \theta_K \) correspond to the bottom model parameters of individual participants, and \( \theta_T \) represents the top model parameters managed by the active party \( P_K \).
During inference, each participant \( P_k \) computes a feature embedding \( e_k = f(X_k; \theta_k) \) by processing its raw feature vector \( X_k \) through its bottom model \( f(\theta_k) \). 
These embeddings are transmitted to \( P_K \), which aggregates them using a function \( \mathtt{Agg}(\cdot) \) (e.g., summation or concatenation) to produce a combined embedding \( E = \mathtt{Agg}(e_1, e_2, \ldots, e_K) \).  

In a conventional VFL system (without defenses), \( E \) is directly input to the top model \( g(\theta_T) \) to generate the final prediction \( \hat{y} = g(E; \theta_T) \), where \( \hat{y} \in [C] \) and \( C \) is the number of classes. 
This prediction is then returned to all passive parties. 
However, practical deployments face risks from malicious passive parties (e.g., \( P_1 \) in Figure \ref{fig:vfl_inference_system}) that may submit adversarial embeddings (e.g., \( e_1 \)) to manipulate predictions via targeted label attacks.  
To mitigate this, the detector-enhanced system introduces an anomaly detector \( \phi(\cdot) \) at \( P_K \). 
After computing \( \hat{y} \), the system evaluates \( \phi(E, \hat{y}) \). 
If \( E \) is flagged as anomalous, the detector triggers a rejection by returning the special label \( \texttt{REJ} \) to all participants; 
otherwise, \( \hat{y} \) is returned as usual. 
For consistency, we treat \( \texttt{REJ} \) as an additional class, extending the total number of classes to \( C+1 \).  

In summary, the detector-enhanced VFL system's prediction is denoted as \( \hat{y} = F(\boldsymbol{X}; \boldsymbol{\theta}) \), where \( \boldsymbol{X} = (X_1, X_2, \ldots, X_K) \), \( F(\cdot) \) combines the VFL model and the detector \( \phi(\cdot) \), and \( \hat{y} \in [C+1] \) includes both the original classes and the \( \texttt{REJ} \) label.

\subsection{Anomaly Detection}  
Anomaly detection (AD) refers to techniques for identifying data patterns that significantly deviate from expected normal behavior. 
These methods fall into two broad categories: non-parametric and parametric approaches. 
This section details two prominent AD techniques: the non-parametric kernel density estimation (KDE) \cite{chen2017tutorial} and the parametric deep autoencoder (DeepAE) \cite{deepae}.  

\textbf{Kernel Density Estimation.}  
KDE is a non-parametric method that estimates the probability density of data instances. 
Let \( X = \{x_1, x_2, \ldots, x_n\} \) denotes $n$ independent and identically distributed observations in $\mathbb{R}^d$, where $d$ is the feature dimension.
For any test instance \( x \in \mathbb{R}^d \), the KDE-based detector computes:  

\begin{equation}  
    \phi(x) = \frac{1}{nh}\sum_{i=1}^n \mathcal{K}\left(\frac{\|x - x_i\|}{h}\right),  
    \label{eq:kde}
\end{equation}  

\noindent where \( h > 0 \) is the bandwidth parameter controlling smoothing intensity, and \( \mathcal{K}(\cdot) \) is a kernel function. 
A common choice is the radial basis function (RBF) kernel: \( \mathcal{K}(z) = \exp(z^2) \), where \( \|x - x_i\| \) represents the Euclidean distance between \( x \) and \( x_i \). 
Eq.(\ref{eq:kde}) operates by constructing localized density contributions around each observation \( x_i \), then aggregating these contributions to estimate the density of \( x \). 

\textbf{Deep Autoencoder.} 
DeepAE is a deep-learning-based parametric AD method that consisting of two neural networks: an encoder \( E(\theta_e) \) that compresses input data \( x \in \mathbb{R}^d \) into a low-dimensional latent vector, and a decoder \( D(\theta_d) \) that reconstructs the input as \( x' \in \mathbb{R}^d \) from this latent representation. 
Here, \( \theta_e \) and \( \theta_d \) denote the trainable parameters of the encoder and decoder, respectively.  
DeepAE models are trained on the observation set \( X \) in an unsupervised learning manner, using the loss function $\mathcal{L} = \frac{1}{n} \sum_{i=1}^{n} \|x_i - x'_i\|^2$, where \( x'_i = D(E(x_i; \theta_e); \theta_d) \). 
After training, for a test instance $x$, DeepAE computes:  
\begin{equation}  
    \phi(x) = \|x - D(E(x; \theta_e); \theta_d)\|^2,  
\end{equation}  
\noindent which quantifies the reconstruction error between the input \( x \) and its decoded output \( x' \). 
Higher scores indicate greater deviation from normal patterns and a higher likelihood of anomaly.

\textbf{Anomaly Score.}
For both KDE and DeepAE, an instance $x$ is detected as an anomaly if its \emph{anomaly score} $\phi(x)$ exceeds a predefined threshold $\tau$:

\begin{equation}
    \phi(x) \ge \tau.
\end{equation}


\textbf{Label-Aware Detection.}
In supervised settings where each observation \( x_i \) is associated with a label \( y_i \in [C] \) (e.g.,  \( C \)-class classification), more accurate anomaly detection can be achieved by utilizing label-specific sub-detectors. 
The detector \( \phi(\cdot) \) is then composed of a set of sub-detectors \( \{\phi_1, \phi_2, \dots, \phi_C\} \), where each sub-detector \( \phi_c \) is trained on observations from \( X \) corresponding to label \( c \). 
For an input instance \( x \) with true label \( y \), 
the anomaly score \( \phi_y(x) \) in computed using the $y$-specific sub-detector.
This label-aware approach ensures higher detection accuracy by leveraging class-specific data distributions.



\subsection{Targeted Label Attack}
A targeted label attack during the inference phase is a type of adversarial attack where an attacker introduces imperceptible perturbations to an original sample.
The goal is to mislead the trained model into classifying the perturbed sample as a specific, targeted label. 

Formally, let \( f(\theta) \) be a well-trained deep learning model, and let \( x \) denote an non-target test sample. 
The attacker's objective can be formulated as:

\begin{equation}
    x' = x + \delta \quad \text{s.t.} \quad f(x';\theta) = y^*, \; f(x';\theta) \neq f(x;\theta), \; \text{and} \; L_p(\delta) \le \epsilon,
\end{equation}

\noindent where \( \delta \) is the perturbation added to the original sample \( x \), \( y^* \) is the targeted label chosen by the attacker, \( f(x; \theta) \) represents the model's prediction for input \( x \), \( L_p(\delta) \) denotes the \( L_p \)-norm, quantifying the perturbation magnitude, and \( \epsilon \) is an upper bound on \( \delta \), ensuring that it remains imperceptible to humans.

The attacker's goal is to craft a perturbation \( \delta \) that satisfies these constraints while minimizing its magnitude. 
This forces the model's prediction to shift from the original label \( f(x; \theta) \) to the targeted label \( y^* \). 
By carefully designing such a perturbation $\delta$ within the defined limit $\epsilon$, the attacker ensures that $x'$ remains visually indistinguishable from $x$ while successfully misleading the model's classification.

\section{Problem Definition}
\subsection{Threat Model} \label{sec:threat_model}

\textbf{Attacker.}
This work focuses exclusively on targeted label attacks occurring during the inference phase, explicitly excluding any attacks or adversarial behaviors during the training phase. 
We assume that all participants adhere honestly to the VFL training protocol for building accurate and reliable VFL model.
However, during the inference phase, we consider that a passive party may be compromised by an attacker with malicious intent to perform a targeted label attack. 
The attacker's goals, knowledge, and capabilities are defined as follows:

\begin{itemize}
    \item \emph{Attacker's goal.} The attacker seeks to deceive the trained VFL model into predicting a specific, targeted label for any given test sample by injecting maliciously crafted features during the inference phase. 
    This attack is well-motivated in real-world applications. 
    For example, in a credit risk assessment scenario, the attacker has strong incentives to manipulate the VFL model's prediction for specific test samples (e.g., corresponding to individual users) from ``high risk'' to ``low risk'' to obtain loans from a financial institution.
    
    \item \emph{Attacker's knowledge.} The attacker operates with minimal prior knowledge, consistent with the standard setup of a VFL inference system.
    Specifically, the attacker knows VFL task type (e.g., classification or regression) and the total number of classes. 
    However, the attacker does not have access to the local datasets of other participants (e.g., their size or distribution) or to the architecture and parameters of their bottom models. 
    Unlike prior works that assume stronger (and often unrealistic) adversarial knowledge \cite{gu2023lrba,qiu2024hijack,pang2023adi,he2023backdoor}, we assume that the attacker has no access to labeled auxiliary samples from either the training or test sets, and cannot access the outputs of the active party's top model (such as logits or probabilities). 
    Additionally, the attacker is unaware of the type of detector deployed by the active party (e.g., KDE or DeepAE). 
    This restricted adversarial knowledge assumption reflects a more realistic threat model, making our approach more applicable to practical scenarios compared to previous works.
    
    \item \emph{Attacker's capability.} The attacker is capable of collecting the predicted label for each test sample. 
    If the predicted label is not the special label $\texttt{REJ}$, the attacker can treat it as a pseudo-label for the corresponding test sample.
    Furthermore, the attacker can manipulate their own feature embeddings or raw feature vectors, but these manipulations are constrained to ensure that the modified features remain within valid ranges.
    For example, in an image dataset, each pixel value in a manipulated image must lie within the range [0, 255]. 
    However, the attacker cannot alter the feature embeddings, raw feature vectors, or bottom model parameters of other participants. 
    Their influence is strictly limited to their own contributions within the VFL system.
\end{itemize}

\textbf{Defender.}
In this work, we also consider the threat model for the defender (the active party).
\emph{Goal:} The defender's objective is to accurately detect and filter out potentially malicious feature embeddings contributed by a compromised passive party during the VFL inference phase. 
The defender then returns the special $\texttt{REJ}$ label to all participants to indicate the presence of malicious activity.
\emph{Knowledge:} The defender is unaware of which passive party has been compromised by the attacker. 
Furthermore, the defender does not have access to the attacker's private test set or to the structure and parameters of the attacker's trained bottom model. 
The defender can only access the feature embeddings received from each passive party.
\emph{Capability:} During the VFL training phase, the defender treats the benign feature embeddings provided by each passive party as observations for constructing anomaly detectors.
In the inference phase, the defender uses these constructed anomaly detectors to evaluate the aggregated feature embeddings and identify whether any of them are malicious.

\subsection{Problem Formulation} \label{sec:problem_definition}
As introduced in Section \ref{sec:vfl_inference_system}, a VFL inference system involves $K$ participants. 
For clarity, these participants are categorized into three types: the attacker $P_{adv}$, the defender (active party) $P_{ap}$, and the remaining normal and honest passive parties $P_{nor}$. 
The corresponding feature vectors provided by these participants are denoted as $X_{adv}$, $X_{ap}$, and $X_{nor}$, respectively.
During the VFL inference phase, the attacker aims to manipulate their feature vector to mislead the VFL model, causing it to predict as many test samples as possible into the targeted label. 
To quantitatively evaluate the effectiveness of the attack, we define the metric \emph{Attack Success Rate (ASR)}.

\begin{definition}[Attack Success Rate]  
    In a detector-enhanced VFL inference system, given the global model $F(\cdot)$ with parameters $\boldsymbol{\theta}$, and a test set consisting of $N$ samples, the attack success rate is defined as:  
    \begin{equation}  
        s(\mathcal{A}) = \frac{\sum_{i=1}^{N} \mathbb{I}(F(\mathcal{A}(X_{adv}^i; \theta_{adv}), X_{nor}^i, X_{ap}^i; \boldsymbol{\theta}) = t^{\star})}{N},  
        \label{eq:asr_definition}  
    \end{equation}  
    where $\theta_{adv}$ represents the attacker's bottom model parameters, $t^{\star}$ is the attacker's targeted label, $\mathcal{A}(\cdot)$ denotes the attack algorithm, and $\mathbb{I}(\cdot)$ is an indicator function that equals 1 if the condition inside is satisfied, and 0 otherwise.  
\end{definition}

Specifically, the attack algorithm $\mathcal{A}(\cdot)$ generates a maliciously crafted sample $X_{adv}^*$ from an original benign sample $X_{adv}$. 
This process is subject to the following constraint:

\begin{equation}  
    X_{adv}^* = \mathcal{A}(X_{adv}; \theta_{adv}) \quad \text{s.t.} \quad dist(X_{adv}^*, X_{adv}) \le r_{max},  
\end{equation}  

\noindent where $dist(\cdot)$ measures the distance between the benign sample $X_{adv}$ and the crafted sample $X_{adv}^*$, and $r_{max}$ is a predefined threshold. 
This constraint ensures:  
(1) the feature values of the crafted sample remain within a valid range, as discussed in Section \ref{sec:threat_model}, and  
(2) the crafted sample avoids detection by anomaly detectors.

As noted in Section \ref{sec:vfl_inference_system}, the output of $F(\cdot)$ includes the special label $\texttt{REJ}$. 
If the prediction of $F(\cdot)$ is $\texttt{REJ}$, which differs from the attacker's targeted label $t^{\star}$, the attack on this test sample is considered unsuccessful. 
Therefore, the objective of this paper is to design an effective attack algorithm $\mathcal{A}$ that requires minimal adversarial knowledge and \emph{maximizes} the ASR in a detector-enhanced VFL inference system.

\section{Observations and Main Idea of VTarbel}
In this section, we conduct exploratory experiments to assess the performance of several representative prior attack methods. 
Our findings reveal that these attacks are largely ineffective when anomaly detectors are employed on the defender's side. 
We further analyze the underlying causes of this limitation and introduce the core ideas behind the design of our attack framework, VTarbel.

\subsection{Observations} \label{sec:observations}
We examine the effectiveness of three previous widely recognized attack methods—LR-BA \cite{gu2023lrba}, HijackVFL \cite{qiu2024hijack}, and ADI \cite{pang2023adi}—in a setting where the defender either has or has not deployed anomaly detectors. 
These methods assume that the attacker can obtain labeled training samples, which are then used to train an auxiliary model. 
The attacker subsequently utilizes optimization algorithms to generate malicious feature vectors for attacking based on this auxiliary model. 
A detailed description of these methods is provided in Section \ref{sec:targeted_output_attack}.

\begin{figure*}[t!]
	\centering

	\begin{minipage}{0.65\linewidth}
		\centering
		\includegraphics[width=0.9\linewidth, trim=40 355 0 0, clip]{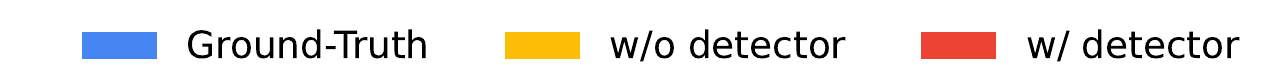}
	\end{minipage}

	\vspace{-2pt}

	\begin{minipage}{1.0\linewidth}
		\centering
		\subfigure[MLP3, TabFMNIST]{
			\label{fig:tabfmnist_observation}
			\includegraphics[width = 0.32\textwidth]{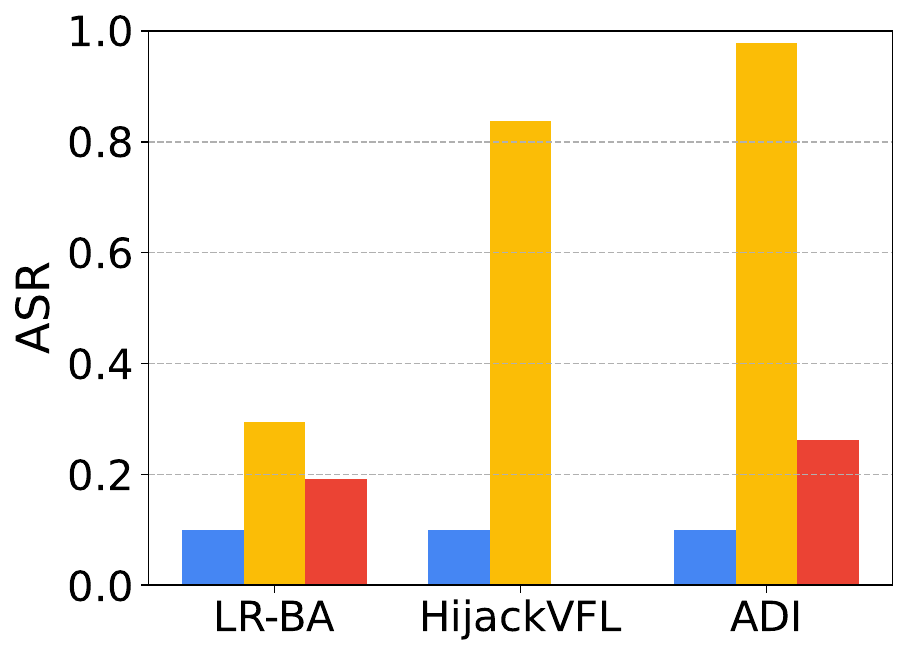}
		}\hfill
		\subfigure[VGG13, CIFAR10]{
			\label{fig:vgg13_cifar10_observation}
			\includegraphics[width = 0.32\textwidth]{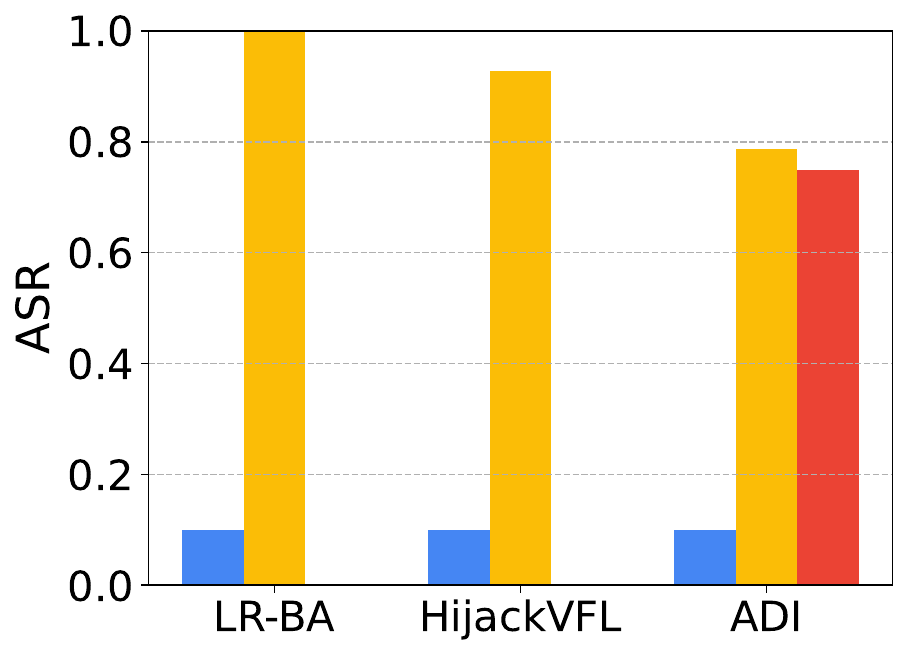}
		}\hfill
		\subfigure[ResNet18, CIFAR10]{
			\label{fig:resnet18_cifar10_observation}
			\includegraphics[width = 0.32\textwidth]{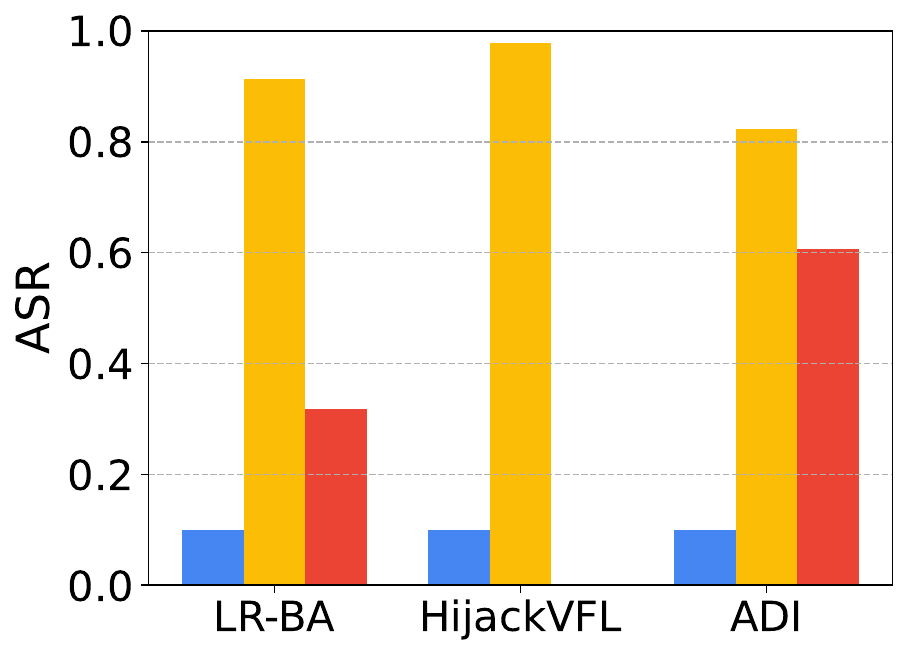}
		}\hfill
	\end{minipage}
	\caption{
		Impact of anomaly detector on ASR in VFL inference system. 
		The ``Ground-Truth'' represents the proportion of the targeted label in the original test set, serving as a baseline for comparison.
	}
	\label{fig:observation}
\end{figure*}

Our experiments are conducted within a two-party VFL  inference system, where the passive party is compromised by the attacker and the active party acts as the defender, employing DeepAE as the anomaly detector. 
We evaluate three different bottom model architectures: MLP3, VGG13 \cite{simonyan2014vgg}, and ResNet18 \cite{he2016resnet}, across two datasets: TabFMNIST \cite{fmnist_dataset} and CIFAR10 \cite{cifar10_dataset}. 
Without loss of generality, we set the targeted label $t^{\star} = 0$ for both datasets. 
The experimental results are presented in Figure \ref{fig:observation}, which reveals two key observations.

\textbf{Observation 1: High ASR without detectors.}
As shown in Figure \ref{fig:observation}, for all three attack methods, the ASR is significantly higher in the absence of anomaly detectors compared to the ``Ground-Truth'' case. 
For instance, in the configuration (MLP3, TabFMNIST), the ASR increases from 10.00\% in the ``Ground-Truth'' case to 29.39\%, 83.68\%, and 97.81\% for LR-BA, HijackVFL, and ADI, respectively. 
These results indicate that the feature vectors, maliciously optimized against the attacker's local auxiliary model, effectively transfer to the global VFL model.

\textbf{Observation 2: Significant drop in attack performance with detectors.}
Figure \ref{fig:observation} also shows a drastic decline in ASR when detectors are deployed by the defender.
In some cases, the ASR even falls to zero, which is even lower than the baseline ``Ground-Truth'' performance. 
For instance, in the (ResNet18, CIFAR10) configuration, the ASR for LR-BA decreases from 91.40\% (without detector) to 31.84\% (with detector). 
This drop is due to the anomaly detector recognizing many malicious feature embeddings as anomalies, causing the global VFL model to predict the special rejection label \(\texttt{REJ}\). 
In other configurations, such as (VGG13, CIFAR10), the ASRs for LR-BA and HijackVFL with detectors are 0.04\% and 0, respectively—significantly lower than the 10.0\% achieved in the ``Ground-Truth'' case. 
This highlights the ineffectiveness of these attacks in practical VFL scenarios where anomaly detectors are deployed.

These results suggest that existing targeted attack methods fail to perform effectively when anomaly detectors are in place to filter out malicious feature embeddings.
This motivates us to develop more robust attack strategies capable of manipulating VFL predictions to target labels while evading detection.

\subsection{Main Idea of VTarbel}
We begin by analyzing why previous attack methods are ineffective. 
While these methods leverage an auxiliary model to optimize the feature vector or embedding, aiming to enhance its predictive capability for the targeted label or reduce the contributions of other benign participants, the malicious inputs generated are \emph{overly} optimized. 
Since these inputs are rarely observed in the training set and fall outside the normal feature space, they have a low probability of occurring and can be easily identified as anomalies by the defender's detector, even under a loose detection threshold. 
As a result, most of these overly optimized inputs are rejected by the global VFL model, leading to the observed low ASR.

To overcome these limitations, we propose VTarbel, a novel two-stage attack framework.
\textbf{The main idea is to partition the VFL inference phase into two substages: the first stage focuses on  collecting essential knowledge, while the second stage executes the actual targeted label attack.}
We elaborate on the main idea below.

\textbf{Leveraging the first stage to estimate detectors and train a surrogate model.}
In the first stage, the attacker honestly follows the VFL inference protocol and collects the returned predicted labels of benign test samples. 
Using these predicted labels, the attacker can:
\emph{(1) estimate the defender's detectors.} 
In ML, it is generally assumed that the training and test datasets share the same distribution. 
As discussed in Section \ref{sec:threat_model}, the defender's detectors are constructed based on the benign feature embeddings from the passive parties during the VFL training phase. 
By leveraging benign feature embeddings from test samples, the attacker can construct approximate detectors (which do not need to be identical to the defender's) to estimate the defender's ground-truth detectors.
\emph{(2) train a surrogate model using local features.}
Inspired by the model completion attack proposed by Fu et al. \cite{fu2022label}, the attacker can append an inference head model (with the same output dimension as the top model) to its well-trained bottom model after VFL training, thereby constructing a surrogate model.
Using the predicted labels and local features, the attacker can build a labeled attack dataset and fine-tune the surrogate model to achieve high accuracy. 
This well-performing surrogate model assists the attacker in generating more effective malicious feature embeddings.

\textbf{Conducting the actual attack in the second stage.}
After the first stage, the defender's detectors are well-estimated, and the surrogate model is well-trained.
In the second stage, the attacker maliciously deviates from the VFL inference protocol and launches the targeted label attack. 
Using the surrogate model and the estimated detectors, the attacker optimizes the raw feature vectors to ensure the surrogate model predicts them as the targeted label with high confidence.
Simultaneously, the attacker ensures that the malicious feature embeddings are not flagged as anomalies by the estimated detectors (e.g., by keeping the anomaly score sufficiently low). 
Once the feature vectors are sufficiently optimized (i.e., the loss is minimized or the iteration limit is reached), the attacker transfers the corresponding embeddings to the active party for VFL inference.

In summary, we propose a novel two-stage attack framework, VTarbel, which will be described in detail in the next section. 
This framework generalizes previous methods, where the ``Ground-Truth'' case and previous attacks illustrated in Figure ~\ref{fig:observation} are special cases of our framework. 
Specifically, the ``Ground-Truth'' corresponds to using only the first stage without performing attacks in the second stage, while previous attacks correspond to cases where only the second stage is applied without the first stage.
In Section ~\ref{sec:stage_ablation}, we demonstrate that both first-stage-only and second-stage-only approaches are ineffective, and only our generalized two-stage framework achieves the highest ASR.

\section{VTarbel Framework Design}

In this section, we first re-formulate the problem definition within the context of the two-stage attack framework (Section \ref{sec:two_stage_formulation}). 
We then provide an overview of the proposed framework, VTarbel (Section \ref{sec:framework_overview}), followed by a detailed explanation of each stage in Sections \ref{sec:preparation_stage} and \ref{sec:attack_stage}.

\subsection{Two-stage Attack Formulation} \label{sec:two_stage_formulation}
In Section \ref{sec:problem_definition}, we define the \emph{attack success rate (ASR)}, \(s(\mathcal{A})\), and present a generalized formulation of the attack objective: maximizing ASR by designing an effective attack algorithm \(\mathcal{A}\). 
In this section, we re-formulate the problem under the two-stage attack framework, providing a more explicit objective function for the attacker.  

For the attacker \(P_{adv}\), it partitions the entire VFL inference phase on its test set \(D_{adv}\) into two substages: a \emph{preparation stage} and an \emph{attack stage}. 
The test samples within the preparation stage, denoted as \(\mathcal{Q} \subset D_{adv}\) with \(\mathcal{Q} \neq \emptyset\), form the \emph{preparation set}. 
The size of this set, or equivalently, the stage length, is \(|\mathcal{Q}|\). 
The remaining test samples, \(D_{adv} \setminus \mathcal{Q}\), constitute the attack stage, with a corresponding stage length of \(N - |\mathcal{Q}|\), where \(N\) is the total number of test samples in \(D_{adv}\).  
Building on the ASR definition in Eq.(\ref{eq:asr_definition}), we define the ASR for the preparation and attack stages as \(s_1\) and \(s_2\), respectively:  

\begin{equation}
    s_1(\mathcal{Q}) = \frac{\sum_{i=1}^{|\mathcal{Q}|} \mathbb{I}(F(X_{adv}^i, X_{nor}^i, X_{ap}^i ; \boldsymbol{\theta}) = t^{\star})}{|\mathcal{Q}|},
\end{equation}

\begin{equation}
    s_2(\mathcal{Q}, \mathcal{A}) = \frac{\sum_{i=|\mathcal{Q}|+1}^N \mathbb{I}(F(\mathcal{A}(X_{adv}^i; \theta_{adv}), X_{nor}^i, X_{ap}^i ; \boldsymbol{\theta}) = t^{\star})}{N - |\mathcal{Q}|}.
\end{equation}

\noindent The attacker's objective can then be formulated as the following optimization problem:  

\begin{equation}
    \mathcal{Q}^* = \argmax_{\mathcal{Q} \subset D_{adv}, \mathcal{Q} \neq \emptyset} \quad \underbrace{s_1(\mathcal{Q}) \cdot \mathcal{|Q|}}_{\text{preparation stage}} +  \underbrace{s_2(\mathcal{Q}, \mathcal{A}) \cdot (N - \mathcal{|Q|})}_{\text{attack stage}}.
    \label{eq:combinatorial_problem}
\end{equation}

\textbf{Main challenge.} 
Directly solving the combinatorial optimization problem in Eq.(\ref{eq:combinatorial_problem}), i.e., finding the optimal preparation set \(\mathcal{Q}^*\) to maximize the number of test samples predicted as the targeted label, is NP-hard \cite{korte2011combinatorial}. 
However, through empirical experiments, we have gained insights into the impact of the preparation stage ratio on both substages. 
The \emph{preparation stage ratio}, denoted as \(\rho = \frac{|\mathcal{Q}|}{N}\), represents the proportion of test samples randomly selected from \(D_{adv}\) for the preparation stage.

\begin{figure*}[t!]
	\begin{minipage}[t]{0.485\linewidth}
		\centering
        \includegraphics[width = \linewidth]{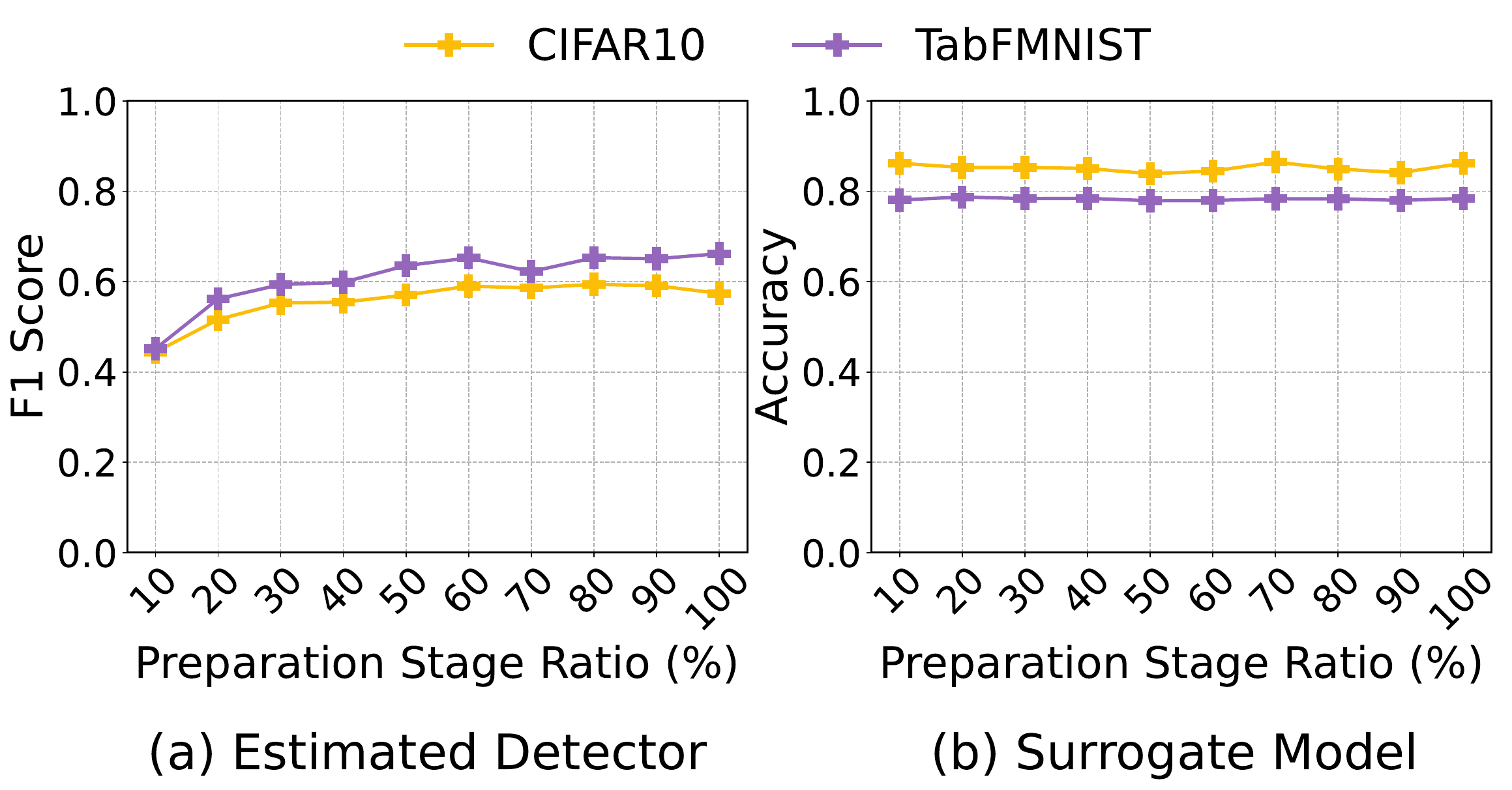}
        \caption{
            Impact of the preparation stage ratio $\rho$ on the performance of attacker's local estimated detector and surrogate model.
        }
		\label{fig:motivation_surrogate_acc_detector_f1}
	\end{minipage}
	\hfill
	\begin{minipage}[t]{0.485\linewidth}
		\centering
        \includegraphics[width = \linewidth]{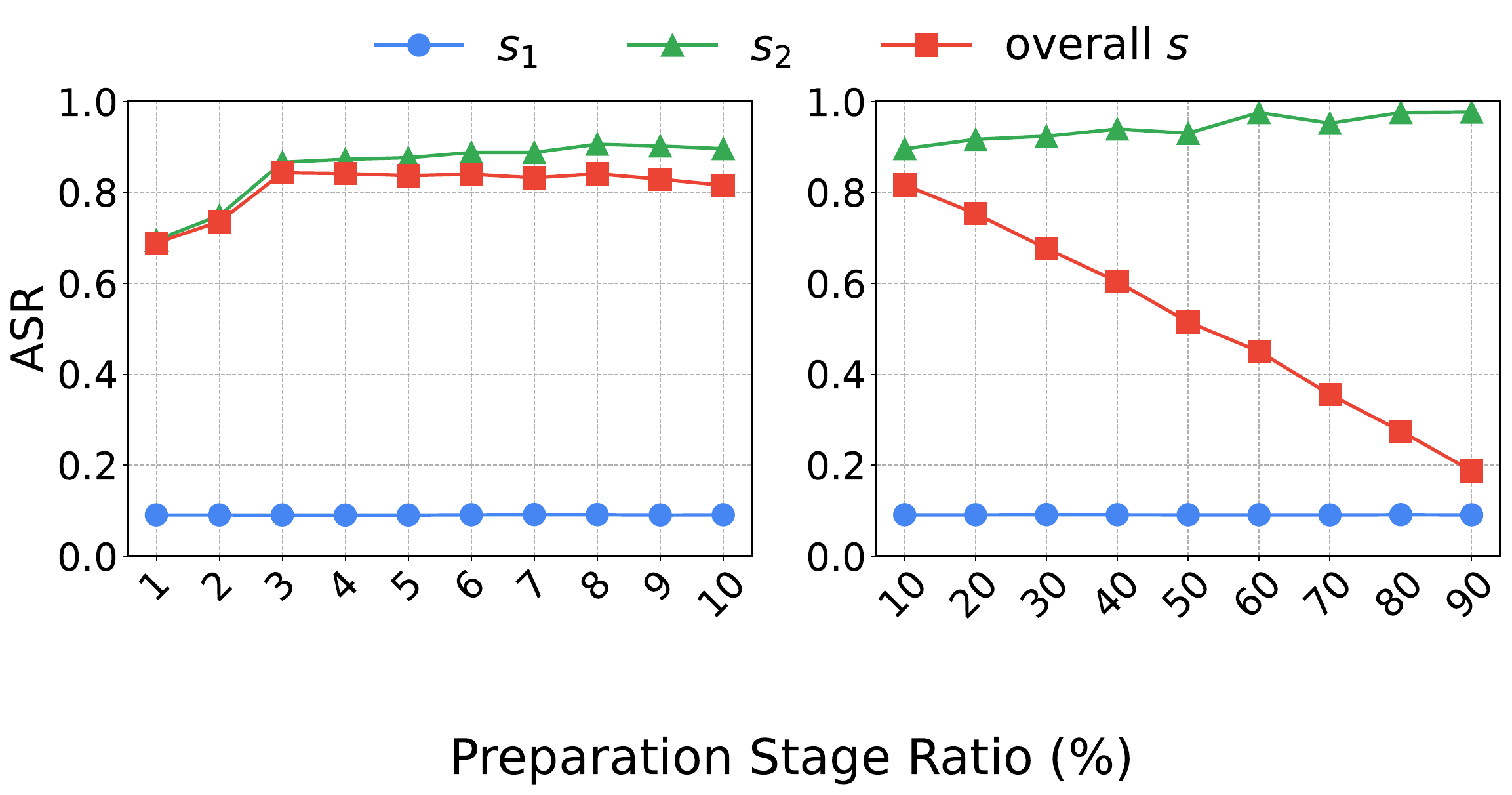}
		\caption{
            Impact of the preparation stage ratio $\rho$ on ASR of each stage and the overall ASR.
        }
		\label{fig:motivation_s1_s2_overall}
	\end{minipage}
    \label{fig:motivation_length_of_preparation}
\end{figure*}

\textbf{Insight 1: Assigning an appropriate number of test samples to \(\mathcal{Q}\) improves the accuracy of the estimated detector and surrogate model, but the marginal benefits diminish as more samples are added.}
We conduct experiments on CIFAR10 and TabFMNIST datasets to evaluate the impact of the preparation stage ratio \(\rho\) on the performance of the estimated detector and surrogate model. 
As shown in Figure \ref{fig:motivation_surrogate_acc_detector_f1}, increasing \(\rho\) from 10\% to 30\% significantly improves the detector's F1 score, with absolute increases of 0.109 for CIFAR10 and 0.142 for TabFMNIST. 
However, when \(\rho\) increases from 30\% to 100\%, the improvements are much smaller, with F1 score increases of only 0.021 and 0.068, respectively. 
For the surrogate model, we find that its accuracy does not improve significantly as \(\rho\) increases. 
When \(\rho = 10\%\), the model accuracies for CIFAR10 and TabFMNIST are 0.8622 and 0.7807, respectively. 
However, when \(\rho\) is increased to 100\%, the corresponding accuracies are 0.8624 and 0.7844, showing negligible improvement.
Based on these results, we conclude that constructing a preparation stage of appropriate length is crucial for accurately estimating detector and training surrogate model. 
However, once sufficient accuracy is reached, further increasing the preparation set size yields only marginal gains.

\textbf{Insight 2: There is trade-off between the preparation stage length and the attack stage length.}
We investigate the impact of \(\rho\) on \(s_1\), \(s_2\), and the overall ASR \(s\) using the ResNet18 model, CIFAR10 dataset, and DeepAE detector.
As shown in Figure \ref{fig:motivation_s1_s2_overall}, we observe that \(s_1\) remains stable and constant with respect to the preparation stage length, while \(s_2\) gradually improves as \(\rho\) increases. 
The stability of \(s_1\) can be attributed to the fact that during the preparation stage, the attacker honestly follows the VFL inference protocol without manipulation. 
Consequently, \(s_1\) is solely dependent on the trained VFL model and is unaffected by the preparation stage length.
In contrast, \(s_2\) improves as \(\rho\) increases. 
As indicated in Figure \ref{fig:motivation_surrogate_acc_detector_f1}, the estimated detector and surrogate model become more accurate with a larger \(\rho\), leading to more effective optimization of feature vectors. 
These optimized features exhibit stronger predictive capability for the targeted label while avoiding detection, thus resulting in a gradual improvement in \(s_2\).

From the above analysis, we identify a fundamental trade-off in the design of the two-stage attack framework. 
From Eq.(\ref{eq:combinatorial_problem}), it is evident that the total number of test samples predicted as targeted label is the summation of two terms: the number of samples predicted as targeted label in  preparation stage and the number predicted in attack stage.
If the preparation stage is too short, although a large number of test samples remain for the attack stage, the relatively low ASR \(s_2\) will result in a small second term in Eq.(\ref{eq:combinatorial_problem}), leading to an ineffective attack.
All previous attack methods fall into this category, corresponding to the special edge case where the preparation stage is empty, i.e., \(\mathcal{Q} = \emptyset\).
On the other hand, if the preparation stage is too long, the ASR \(s_2\) will be high, but the number of test samples remaining for the attack stage will be small. 
This also causes the second term in Eq.(\ref{eq:combinatorial_problem}) to be small, resulting in an ineffective attack. 
The ``Ground-Truth'' case in Figure ~\ref{fig:observation} falls into this category, corresponding to the special edge case where \(\mathcal{Q} = D_{adv}\).

The experimental results in Figure ~\ref{fig:motivation_s1_s2_overall} confirm our analysis. 
The overall ASR \(s\) initially improves as \(\rho\) increases from 1\% to 5\%. 
However, when \(\rho\) becomes relatively large (10\% to 90\%), there is a corresponding decrease in \(s\).
To effectively address this trade-off, our design goal is to construct a preparation stage that maximizes \(s_2\) while keeping its length minimal, thereby preserving as many test samples as possible for the attack stage.

\subsection{Framework Overview} \label{sec:framework_overview}

\begin{figure*}[t]
    \centering
    \includegraphics[width=1.0\linewidth]{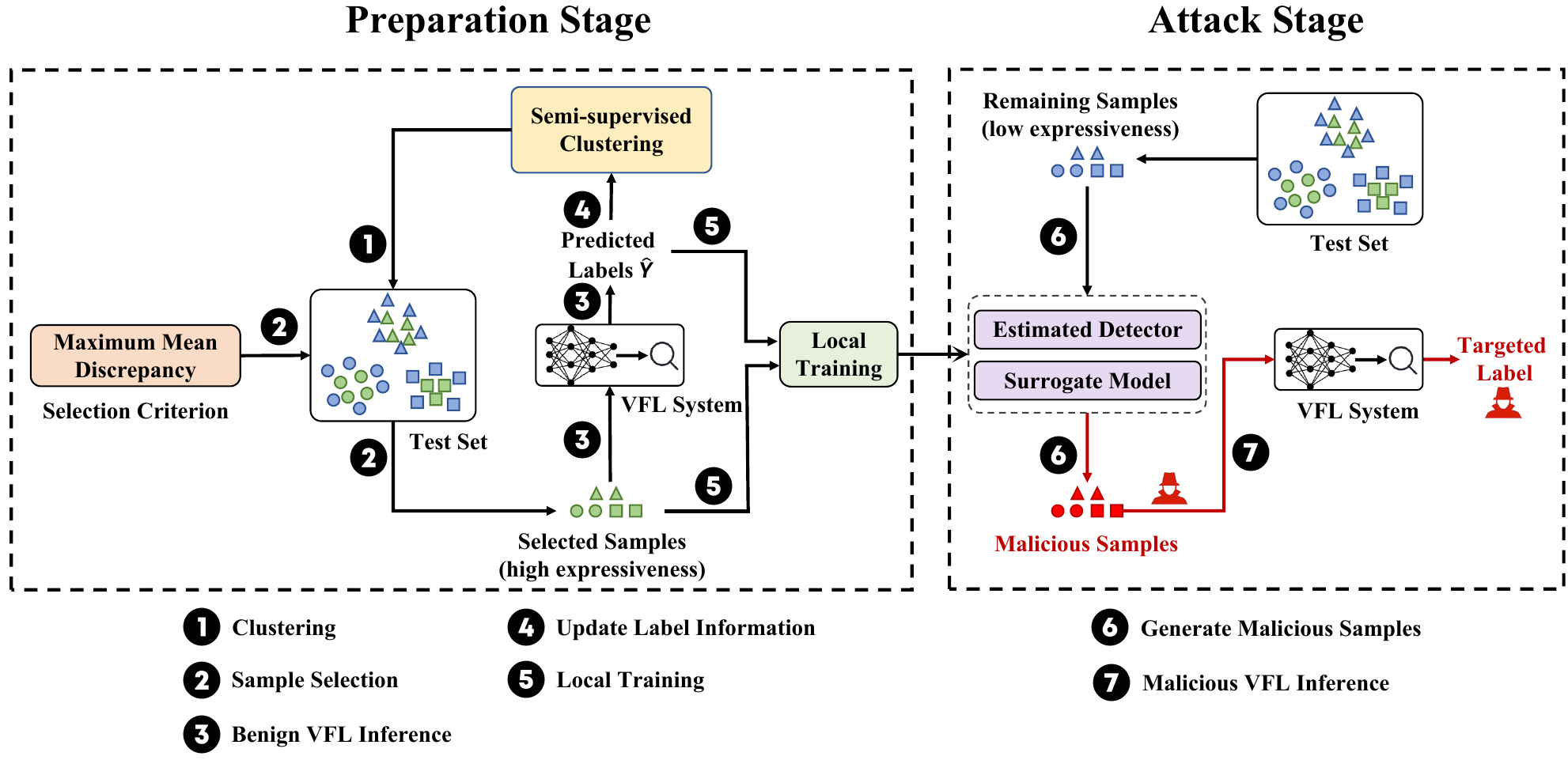}
    \caption{
        Overview of the two-stage attack framework, VTarbel. 
        The \textcolor{green}{green} samples represent test samples with high expressiveness, while the \textcolor{blue}{blue} samples represent those with lower expressiveness. 
        The \textcolor{red}{red} samples denote the maliciously generated samples fed into the VFL inference system. 
        Different shapes (circle, triangle, and square) indicate test samples from different classes.
    }
    \label{fig:vtarble_framework}
\end{figure*}

Figure \ref{fig:vtarble_framework} illustrates the overview of our two-stage attack framework, VTarbel, which splits the VFL inference phase on the entire test set into a preparation stage and an attack stage.

\textbf{In the preparation stage}, the attacker follows these steps to gather essential knowledge. 
(1) The attacker initially applies a semi-supervised clustering technique (e.g., constrained seed K-Means \cite{basu2002semi}) to partition the unlabeled test set into \(C\) clusters.
(2) Within each cluster, the attacker selects test samples with high expressiveness. 
Expressiveness is measured by the ability to maximize the reduction in the maximum mean discrepancy (MMD) ~\cite{mmd_definition}, a metric that quantifies the distribution divergence between the selected samples and the full test set.
(3) The selected expressive samples are then fed into the VFL inference system to obtain the predicted labels \(\hat{Y}\).
(4) Using the newly predicted labels, the semi-supervised clustering algorithm is updated to improve the accuracy of the clusters.
These steps are iterated until the MMD falls below a predefined threshold or the maximum number of unlabeled sample selection rounds is reached.
(5) The selected samples and their corresponding predicted labels are collected as pairs, which are then used for locally training the estimated detectors and the surrogate model.

\textbf{In the attack stage}, the attacker uses the estimated detector and surrogate model trained in the previous stage to execute the attack.
(6) The attacker filters the remaining test samples (those with lower expressiveness) and feeds them into both the surrogate model and the detector for local gradient-based optimization. 
The optimization objective is to maximize the probability that the surrogate model predicts the optimized samples as the targeted label, while avoiding detection by the estimated detector.
(7) Once the optimization is complete, the generated malicious samples are transferred to the VFL inference system. 
The goal is for the global VFL model to predict these samples as the targeted label \(t^{\star}\), regardless of the benign samples provided by other participants.


\subsection{Preparation Stage} \label{sec:preparation_stage}
As shown in Algorithm ~\ref{algo:preparation}, the goal of the preparation stage in VTarbel is to select as few unlabeled test samples with high expressiveness as possible to construct \(\mathcal{Q}^*\), and to train an accurate estimated detector \(\phi_{est}\) and surrogate model \(f_{sur}(\theta_{sur})\). 
This enables the attacker to achieve a high ASR \(s_2\) while preserving as many test samples as possible for the attack stage.

\underline{\textcircled{0} Initialization.}
The attacker begins by initializing the preparation set \(\mathcal{Q}^* = \emptyset\). 
For each class \(c \in [C]\), a labeled sample set \(\mathcal{Q}_c = \emptyset\) is also initialized to track predicted (labeled) samples (Line 1).  
Next, the attacker locally constructs an estimated detector \(\phi_{est}\) and a surrogate model \(f_{sur}(\theta_{sur}) = f(\theta_{adv}) \circ f(\theta_h)\), where \(f(\theta_{adv})\) is the attacker's trained bottom model from the VFL training phase, and \(f(\theta_h)\) is a randomly initialized inference head. 
Here, \(\circ\) denotes model concatenation (Lines 2-3).  
The sample selection round is initialized as \(t \gets 1\), the initial MMD score as \(mmd_0 \gets \infty\), and the locally collected labeled dataset as \(D_{loc} = \emptyset\) (Line 4).

\underline{\textcircled{1} Semi-supervised clustering.}
To construct a balanced preparation set, ensuring an approximately equal number of test samples per class for faster training of estimated detectors, the attacker employs a clustering-based approach to partition the unlabeled test set into \(C\) clusters. 
An equal number of samples \(\eta\) is then selected from each cluster.  
Semi-supervised clustering is used instead of vanilla unsupervised clustering because the predicted labels returned during inference can serve as auxiliary supervision, improving clustering quality. 
Specifically, the attacker adopts the constrained seed K-Means algorithm \cite{basu2002semi}, which is known for its simplicity and superior performance over standard K-Means \cite{krishna1999genetic}.  
For further details on this algorithm, refer to Figure 2 in \cite{basu2002semi}. 
In Line 6, the \texttt{ConstrainedSeedKMeans} function takes three inputs: the unlabeled test set $D_{adv}$, a subset of labeled samples $D_{loc}$, and the desired number of clusters \(C\), returning clusters \(S_1, S_2, \dots, S_C\).

\begin{algorithm}[t]
    \small
    \caption{Preparation Stage in VTarbel}
    \label{algo:preparation}
    \SetKwInOut{Input}{Input}
    \SetKwInOut{Output}{Output}
    \Input{
        Test dataset $D_{adv}$, trained global VFL model $F(\boldsymbol{\theta})$, total number of classes $C$, local fine-tuning epochs $T_{ft}$, test sample selection step $\eta$, maximum selection rounds $R$, MMD stopping tolerance $\epsilon$.
    }
    \Output{
        Test samples in preparation stage $\mathcal{Q}^*$, estimated detector $\phi_{est}$, surrogate model $f_{sur}(\theta_{sur})$.
    }

    \tcc{\textcircled{0} Initialization}
    Initialize $\mathcal{Q}^* = \emptyset$, for each class $c\in[C]$, initialize its corresponding labeled sample set $\mathcal{Q}_c = \emptyset$ \;
    Construct a randomly initialized estimated detector $\phi_{est}$ \;
    Randomly initialize an inference head model $f(\theta_h)$, append it to the trained bottom model to construct the surrogate model $f_{sur}(\theta_{sur}) = f(\theta_{adv}) \circ f(\theta_h)$ \;
    Set selection round $t \gets 1$, initial MMD score $mmd_0 \gets \infty$, and collected labeled sample pairs $D_{loc} = \emptyset$ \;

    \Repeat{
        $\left|mmd_t - mmd_{t-1}\right| \le \epsilon$ or $t > R$
    }{
        \tcc{\textcircled{1} Semi-supervised clustering}
        $S_1, S_2, \cdots, S_C \gets $ \texttt{ConstrainedSeedKMeans}$(D_{adv}, D_{loc}, C)$ \;

        \tcc{\textcircled{2} Sample selection}
        Set candidate sample set for current round $Z = \emptyset$ \;
        \For{each cluster $S_c, c\in[1, C]$}{
            Set sample expressiveness set $E_c = \emptyset$ \;
            \For{each unlabeled sample $x_{u} \in S_c \setminus \mathcal{Q}_c$}{
                Calculate sample expressiveness $e_{u}$ according to Eq.(\ref{eq:expressiveness}), add $e_{u}$ to $E_c$ \;
            }
            Sort $E_c$ in descending order, select the first $\eta$ samples $S_{c,\eta}$ \;
            Update $Z \gets Z \cup S_{c,\eta}, \mathcal{Q}_c \gets \mathcal{Q}_c \cup S_{c,\eta}, \mathcal{Q}^* \gets \mathcal{Q}^* \cup S_{c,\eta}$ \;
        }

        \tcc{\textcircled{3} Benign VFL inference}
        \For{each selected sample $x_{z}\in Z$}{
            Feed $x_z$ into global VFL model to get $\hat{y}_z = F(x_z; \boldsymbol{\theta})$ \;
            \tcc{\textcircled{4} Update label information for clustering}
            Add pair $(x_z, \hat{y}_z)$ into $D_{loc}$ \;
        }
        Calculate $mmd_t \gets \text{MMD}^2(D_{adv}, \mathcal{Q}^*)$\;
        $t \gets t + 1$ \;
    }
    \tcc{\textcircled{5} Local training}
    Fine-tune $f_{sur}$ with $D_{loc}$ in $T_{ft}$ epochs, update $\phi_{est}$ with $D_{loc}$ \;
    

   \Return{$\mathcal{Q}^*$, $\phi_{est}$, $f_{sur}(\theta_{sur})$.} 
\end{algorithm}

\underline{\textcircled{2} Sample selection.}
The core component of the preparation stage lies in the sample selection step. 
In each selection round, the attacker initializes a candidate set \(Z = \emptyset\) to store the selected samples for the current round (Line 7).  
For each cluster \(S_c\), where \(c \in [1, C]\), the attacker iterates over the unlabeled samples \(x_u \in S_c \setminus \mathcal{Q}_c\), computes the expressiveness \(e_u\) of each sample, and stores it in the corresponding expressiveness set \(E_c\) (Lines 8-11).

In this paper, we adopt maximum mean discrepancy (MMD) \cite{mmd_definition} as the foundation for computing sample expressiveness. 
MMD measures the distributional divergence between two sets \(X = \{x_1, \ldots, x_m\}\) and \(Y = \{y_1, \ldots, y_n\} \subseteq X\), and is defined as:

\begin{equation}
    \text{MMD}^2 (X, Y) = \frac{1}{m^2}\sum_{i,j}k(x_i, x_j) + \frac{1}{n^2}\sum_{i,j} k(y_i, y_j) - \frac{2}{mn} \sum_{i,j} k(x_i, y_j),
    \label{eq:mmd_definition}
\end{equation}

\noindent where \(m\) and \(n\) are the number of samples in \(X\) and \(Y\), respectively, and \(k(\cdot,\cdot)\) denotes the kernel function (e.g., the RBF kernel). 
A smaller MMD value indicates greater similarity between the two distributions; MMD equals zero if and only if the distributions are identical.

From Figure \ref{fig:motivation_surrogate_acc_detector_f1}, we observe that the preparation set size has minimal impact on the surrogate model's accuracy. 
Therefore, the attacker focuses primarily on accurately estimating the defender's detector. 
This is equivalent to minimizing the distributional discrepancy between the preparation set \(\mathcal{Q}^*\) and the full test set \(D_{adv}\), which directly aligns with the objective to minimize $ \text{MMD}^2(D_{adv}, \mathcal{Q}^*)$.
To this end, for any unlabeled sample \(x_u\), we define its effectiveness \(e_u\) as the reduction in MMD when the sample is added to the current preparation set:

\begin{equation}
    \begin{aligned}
        e_{u} &= \Delta \text{MMD}^2 \\
        &= \text{MMD}^2(D_{adv}, \mathcal{Q}^*) - \text{MMD}^2(D_{adv}, \mathcal{Q}^* \cup \{x_u\}) \\
        &= \frac{1}{(|\mathcal{Q}^*| + 1)^2} \left( k(x_u, x_u) + 2\sum_{y\in \mathcal{Q}^*} k(y, x_u) \right) - \frac{2}{N(|\mathcal{Q}^*| + 1)} \sum_{x\in D_{adv}} k(x, x_u).
    \end{aligned}
    \label{eq:expressiveness}
\end{equation}

\noindent Eq.(\ref{eq:expressiveness}) quantifies how much the MMD is reduced by adding \(x_u\) to \(\mathcal{Q}^*\). 
A larger value indicates that the inclusion of \(x_u\) brings the preparation set distribution closer to that of \(D_{adv}\), which is interpreted as higher expressiveness.


Using Eq.(\ref{eq:expressiveness}), the attacker computes the effectiveness for all candidate samples in each cluster. 
To mitigate the computational overhead of repeatedly evaluating MMD, the attacker can employ an incremental update strategy inspired by online kernel methods \cite{engel2004kernel}, enabling linear-time updates using precomputed statistics rather than recomputing the full kernel matrix.
Then, the expressiveness set \(E_c\) is sorted in descending order, and the top \(\eta\) samples with the highest effectiveness scores are selected to form \(S_{c,\eta}\) (Line 12).  
Finally, the sample sets \(Z\), \(\mathcal{Q}_c\), and \(\mathcal{Q}^*\) are updated to include the selected \(S_{c,\eta}\) (Line 13).

\underline{\textcircled{3} Benign VFL inference.}
For each selected unlabeled sample \(x_z\) in the candidate set \(Z\) during the current selection round, the attacker honestly feeds \(x_z\) into the VFL inference system to obtain the predicted label \(\hat{y}_z = F(x_z; \boldsymbol{\theta})\) (Lines 14-15).

\underline{\textcircled{4} Update label information.}
For each test sample \(x_z \in Z\), if the predicted label \(\hat{y}_z \neq \texttt{REJ}\), the attacker adds the pair \((x_z, \hat{y}_z)\) to the local labeled dataset \(D_{loc}\) (Line 16). 
This updated dataset, now containing more labeled samples, will be used for the \texttt{ConstrainedSeedKMeans} algorithm in the next selection round to achieve more accurate clusters.

Steps \textcircled{1}-\textcircled{4} iterates until either of the following conditions is met: the absolute difference in the MMD score between two consecutive selection rounds falls below the tolerance \(\epsilon\), or the maximum number of selection rounds \(R\) is reached (Line 19).

\underline{\textcircled{5} Local training.}
Using the locally collected labeled dataset \(D_{loc}\), the attacker fine-tunes the surrogate model \(f_{sur}\) for \(T_{ft}\) epochs and updates the local estimated detector \(\phi_{est}\) accordingly (Line 20). 
Specifically, each sub-detector \(\phi_{est}^i\) (\(i \in [C]\)) in \(\phi_{est}\) is updated using the data samples in \(D_{loc}\) with the predicted label \(\hat{y} = c\).





\subsection{Attack Stage} \label{sec:attack_stage}

As illustrated in Algorithm \ref{algo:attack}, the attack stage involves the attacker performing the actual attack by leveraging a gradient-based optimization method to generate malicious samples. 
These samples are crafted to manipulate the VFL inference system into outputting the targeted label.

\underline{\textcircled{6} Generate malicious sample.}
For each unlabeled test sample \(x\) in the remaining set \(D_{adv} \setminus \mathcal{Q}^*\), the attacker generates the corresponding malicious sample \(x_{adv}\) by minimizing the following adversarial objective function (Lines 2-8):

\begin{equation}
    \begin{gathered}
        J(x) = \texttt{CE}(\text{Softmax}(f_{sur}(x)), t^{\star}) + \lambda \phi_{est}(x), \\
        x_{adv} = \argmin_{x} \; J(x), \\
        \text{s.t.} \quad dist(x_{adv}, x) \le r_{max},
    \end{gathered}
    \label{eq:optimization}
\end{equation}

\noindent where \(\texttt{CE}(\cdot)\) denotes the cross-entropy loss function.
The first term of \(J(x)\) is the main objective, representing the cross-entropy loss between the predicted probability of sample \(x\) by the surrogate model \(f_{sur}\) and the targeted label \(t^{\star}\). 
Minimizing this loss ensures that \(f_{sur}\) predicts \(x\) as \(t^{\star}\) with high confidence.
The second term acts as a regularization term, which restricts the optimized sample within the valid region of the normal distribution. 
This prevents the sample from being detected as anomalous by the estimated detector \(\phi_{est}\), thereby increasing the likelihood of evading detection by the defender's detector.
These two terms are balanced by the hyperparameter \(\lambda\). 
This optimization problem is subject to the constraint that the distance between the original sample \(x\) and the malicious sample \(x_{adv}\) does not exceed a predefined tolerance \(r_{max}\), ensuring that \(x_{adv}\) retains its validity within the feature space.

The above constrained optimization problem can be solved using projected gradient descent (PGD) \cite{pgd}. 
Specifically, in the \(t\)-th optimization step, as shown in Eq.(\ref{eq:pgd}), the attacker performs the following updates:

\begin{equation}
    \label{eq:pgd}
    \begin{gathered}
        x'_t \gets x_t - \alpha \nabla_{x_t} J(x_t), \\
        x_{t+1} \gets \texttt{Proj}(x'_t), 
    \end{gathered}
\end{equation}

\noindent where \(x'_t\) is the intermediate sample obtained by performing vanilla stochastic gradient descent (SGD) on \(x_t\), and \(\alpha\) is the learning rate. 
The attacker then applies the \(\texttt{Proj}(\cdot)\) function to project the intermediate sample \(x'_t\) back into the valid region that satisfies the constraint, yielding \(x_{t+1}\) for the next iteration.
This optimization terminates if either the anomaly score from $\phi_{est}$ exceeds the detection threshold \(\tau\) (Lines 7-8), or the maximum number of steps is reached.

\underline{\textcircled{7} Malicious VFL inference.}
After generating the malicious sample \(x_{adv}\) corresponding to each unlabeled sample \(x\) in the remaining set, the attacker feeds \(x_{adv}\) into the VFL inference system (Line 9). 
The attack on sample \(x\) is considered successful if the predicted label \(\hat{y}_{adv}\) equals the targeted label \(t^{\star}\) (Line 10).

The intuition behind this attack strategy is rooted in the transferability property of malicious samples \cite{gu2023survey, inkawhich2019feature}. 
Specifically, if the malicious samples can deceive the surrogate model by causing it to predict the targeted label \(t^{\star}\) with high confidence, while simultaneously avoiding detection by the estimated detector as an anomaly, the malicious sample is likely to transfer its inherent adversarial properties to the global VFL model. 
This increases the likelihood that the global model will also output the targeted label, thereby achieving the adversarial goal.
Our experimental evaluation in Section \ref{sec:attack_effectiveness} confirms this theoretical foundation, demonstrating the practical viability of our attack strategy.

\begin{algorithm}[t]
    \small
    \caption{Attack Stage in VTarbel}
    \label{algo:attack}
    \SetKwInOut{Input}{Input}
    \SetKwInOut{Output}{Output}
    \Input{
        Test dataset $D_{adv}$, trained global VFL model $F(\boldsymbol{\theta})$, test samples in preparation stage $\mathcal{Q}^*$, estimated detector $\phi_{est}$, anomaly detection threshold $\tau$, local surrogate model $f_{sur}(\theta_{sur})$, targeted label $t^{\star}$, maximum optimization steps to generate malicious sample $T_{opt}$.
    }
    
    Filter out remaining samples in the attack stage as $D_{adv} \setminus \mathcal{Q}^*$ \;

    \For{each unlabeled sample $x \in D_{adv} \setminus \mathcal{Q}^*$}{
        \tcc{\textcircled{6} Generate malicious sample}
        Initialize $x_1 \gets x$ \;
        \For{optimization step $t \in [1, T_{opt}]$}{
            Calculate adversarial loss $J(x_t)$ according to Eq.(\ref{eq:optimization}) \;
            Optimize $x_t$ according to Eq.(\ref{eq:pgd}) to obtain $x_{t+1}$ \;
            \If{$\phi_{est}(x_{t+1}) > \tau$}{
                \textbf{break} \;
            }
        }

        \tcc{\textcircled{7} Malicious VFL inference}
        Set $x_{adv} \gets x_t$, feed $x_{adv}$ into global VFL model to get $\hat{y}_{adv} = F(x_{adv}; \boldsymbol{\theta})$ \;
        Check whether $\hat{y}_{adv} = t^{\star}$ to determine if the attack on sample $x$ is successful \;
    }

\end{algorithm}


\section{Evaluation}

\subsection{Experimental Setup} \label{exp:setting}
\noindent \textbf{Models and Datasets.} 
We evaluate our attack framework using four different model architectures: a fully-connected multi-layer perceptron (MLP3, 3 means three layers), convolutional-based VGG13 \cite{simonyan2014vgg} and ResNet18 \cite{he2016resnet}, and Transformer-based DistilBERT \cite{sanh2019distilbert} (which has been rarely explored in previous studies). 
These models are used for both the active and passive parties in the VFL setting.
For the top model and the attacker's inference head model, we use a simple MLP2 architecture. 
These bottom models are evaluated using classification datasets of varying modalities, numbers of test samples, and number of classes, as summarized in Table \ref{tab:models_datasets}.
(1) For MLP3, we evaluate it on \emph{tabular} datasets TabMNIST and TabFMNIST, where each data sample is generated by flattening the original \(28 \times 28\) grayscale images from the MNIST \cite{mnist_dataset} and FashionMNIST \cite{fmnist_dataset} datasets into a 784-dimensional feature vector.
(2) For VGG13, We use \emph{image} datasets CIFAR10 \cite{cifar10_dataset} and CINIC10 \cite{cinic10_dataset}. 
Both datasets contain 10 classes, with CINIC10 containing 90,000 samples—significantly more than CIFAR10, which has 10,000 test samples. 
This allows us to assess the attack performance on a larger dataset.
(3) For ResNet18, it is evaluated on the CIFAR10 and CIFAR100 \cite{cifar10_dataset} \emph{image} datasets.
The CIFAR10 dataset is used to compare attack effectiveness across different bottom model architectures on the same dataset, while CIFAR100 (with 100 classes) allows us to evaluate the attack performance on datasets with a much larger number of classes.
(4) For DistilBERT, we leverage \emph{text} datasets commonly used for text classification tasks, including TREC \cite{trec_dataset} and IMDB \cite{imdb_dataset}.
For both image and text datasets, each image or text sample is evenly split into sub-images or sub-texts, respectively. 
Similarly, for tabular datasets, each raw feature vector is also evenly split into sub-vectors for distribution across each party.

\noindent \textbf{Metrics.}
For targeted label attack, we report the \emph{Top1} ASR for all datasets, except for CIFAR100 dataset, where we report \emph{Top5} ASR.
Similarity, for main VFL inference task, we report \emph{Top1} accuracy for all datasets, except for CIFAR100 dataset, where we report \emph{Top5} accuracy.
For the performance of the attacker's local estimated detector, we report the \emph{F1 score}.

\noindent \textbf{Baseline.}
We report the ASR achieved when the passive party honestly follows the inference protocol in a detector-enhanced VFL system, which serves as the attack baseline.

\noindent \textbf{Compared Methods.}
We compare VTarbel with four representative targeted label attacks in VFL: LR-BA \cite{gu2023lrba}, HijackVFL \cite{qiu2024hijack}, ADI \cite{pang2023adi}, and TIFS'23 \cite{he2023backdoor}.
For detailed description of these attack methods, please refer to Section \ref{sec:targeted_output_attack}.

\begin{table}[t]
    \caption{Model architectures, dataset statistics, and configurations of VTarbel.}
    \centering
    \resizebox{\linewidth}{!}{%
    \begin{threeparttable}[t]
    \footnotesize
    \begin{tabular}{cccc||ccccc||cccccc}
        \toprule
        \makecell{Bottom\\Model} & \makecell{Top\\Model} & \makecell{Surrogate\\Inference\\Head} & DeepAE & Dataset  & Modality & \makecell{No.\\Test\\Samples} & \makecell{Embedding \\Dimension} & \makecell{No.\\Classes} & \makecell{$\eta$} & \makecell{$\epsilon$} & \makecell{$T_{ft}$} & \makecell{$\lambda$ for \\ DeepAE} & \makecell{$\lambda$ for \\ KDE} & \makecell{$T_{opt}$}\\
        \midrule
        \multirow{2}{*}{MLP3} & \multirow{2}{*}{MLP2} & \multirow{2}{*}{MLP2} & \multirow{2}{*}{MLP6} & TabMNIST & \multirow{2}{*}{Tabular} & 10,000 & 64 & 10 & 10 & 1e-4 & 50 & 1 & 20 & 50 \\
        & & & & TabFMNIST & & 10,000 & 64 & 10 & 10 & 1e-4 & 50 & 1 & 20 & 50 \\
        \cmidrule(lr){1-15}

        \multirow{2}{*}{VGG13} & \multirow{2}{*}{MLP2} & \multirow{2}{*}{MLP2} & \multirow{2}{*}{MLP6} & CIFAR10  & \multirow{2}{*}{Image} & 10,000 & 10 & 10 & 10 & 1e-4 & 50 & 1 & 20 & 50\\
        & & & & CINIC10  & & 90,000 & 10 & 10 & 50 & 1e-4 & 50 & 1 & 20& 50\\
        \cmidrule(lr){1-15}

        \multirow{2}{*}{ResNet18} & \multirow{2}{*}{MLP2} & \multirow{2}{*}{MLP2} & \multirow{2}{*}{MLP6} & CIFAR10  & \multirow{2}{*}{Image} & 10,000 & 10 & 10 & 10 & 1e-4 & 50 & 1 & 20  & 50\\
        & & & & CIFAR100  & &  10,000 & 100 & 100 & 5 & 1e-5 & 50 & 1 & 20& 50\\
        \cmidrule(lr){1-15}

        \multirow{2}{*}{DistilBERT} & \multirow{2}{*}{MLP2} & \multirow{2}{*}{MLP2} & \multirow{2}{*}{MLP6} & TREC & \multirow{2}{*}{Text} & 554 & 64 & 6 & 5 & 1e-2 & 20 & 1 & 20  & 50\\
        & & & & IMDB  & & 25,000 & 64 & 2 & 100 & 1e-3 & 20 & 1 & 20 & 50 \\
        
        \bottomrule
    \end{tabular}

    \begin{tablenotes}[flushleft]
        \footnotesize
        \item $\eta$: sample selection step, $\epsilon$: MMD stopping tolerance, $T_{ft}$: local fine-tuning epochs, $\lambda$: balancing parameter, $T_{opt}$: maximum optimization steps. 
    \end{tablenotes}
    \end{threeparttable}
    }
    \label{tab:models_datasets}
\end{table}

\noindent \textbf{Hyperparameters.}
(1) For VTarbel, the hyperparameters for each dataset are listed in Table \ref{tab:models_datasets}. 
These include parameters in the preparation stage, such as the test sample selection step \(\eta\), MMD stopping tolerance \(\epsilon\), and local fine-tuning epochs \(T_{ft}\), as well as parameters in the attack stage, such as the balancing parameter \(\lambda\) and the maximum number of optimization steps \(T_{opt}\).
We calculate the MMD value in Eq.(\ref{eq:expressiveness}) in the feature embedding space, rather than in the raw feature vector space.
(2) For LR-BA, the auxiliary labels for each class are set to 4 across all datasets, and the number of optimization iterations for generating the backdoor latent representation is set to 50.
(3) For HijackVFL, we follow the settings from the original paper, where the backdoor poison ratio of the target class is set to 1\%, the number of available labels for each class is set to 4, the total epochs for training the surrogate model is set to 50, and the standard variance for generating Gaussian noise is set to 0.1.
(4) For ADI, the attacker is granted maximum adversarial knowledge, meaning it knows the parameters and outputs of the top model and has direct access to the defender's raw feature vectors. 
The optimization rounds for generating the adversarial dominant inputs are set to 50.
(5) For TIFS'23, in accordance with the original paper, during VFL training, 10\% of the training samples from the targeted label are assigned to the attacker. 
During the inference phase, the top 50 samples predicted with the highest probability for the targeted label are selected to generate the trigger embedding.

Unless otherwise specified, the experiments are conducted in a two-party VFL inference system, where the passive party is compromised by the attacker and the active party is the defender.
The defender's anomaly detection threshold is set to the 95\% percentile of the anomaly scores computed from the training samples, and the default anomaly detector used by the defender is DeepAE.
The targeted label is set to ``0'' for all datasets, except for TREC, where it is set to ``1''.
The impact of the detection threshold and the targeted label is discussed in Section \ref{sec:ablation_study}.


\noindent \textbf{Implementation.}
We built a VFL system on a cluster of Ubuntu 22.04 servers, each equipped with a 64-core Intel(R) Xeon(R) Gold 5420+ CPU, 128GB RAM, and two NVIDIA A40 GPUs. 
The servers communicate with each other via Gigabit Ethernet.  
For training, inference, and malicious sample generation with MLP and convolutional models, we use the PyTorch framework \cite{NEURIPS2019_9015}. 
For the Transformer-based model, we adopt the HuggingFace framework \cite{wolf-etal-2020-transformers}. 
Communication among parties is managed using the Python socket library \cite{python-socket}.  
For attack methods with open-source code, we use their available implementations. 
Otherwise, we reproduce the methods ourselves.
Each attack method is evaluated under three independent VFL model trainings, and the reported ASRs are averages from these experiments.

\begin{table}
    \footnotesize
    \centering
    \caption{Comparison of attack performance (ASR (\%)) of various attack methods. 
    }
    \resizebox{\linewidth}{!}{%

    \begin{threeparttable}[t]
    \footnotesize
    \begin{tabular}{c||c||cc|cc|cc|cc}
        \toprule
        \multirow{2}{*}{\makecell{Defender's \\ Detector}}  & \multirow{2}{*}{Method}  &  \multicolumn{2}{c|}{MLP3} & \multicolumn{2}{c|}{VGG13} & \multicolumn{2}{c|}{ResNet18} & \multicolumn{2}{c}{DistilBERT} \\
        & & TabMNIST & TabFMNIST & CIFAR10 & CINIC10 & CIFAR10 & CIFAR100 & TREC & IMDB \\
        \midrule
        — & Ground-Truth & 9.80 & 10.00 & 10.00 & 10.00 & 10.00 & 5.00 & 16.90 & 50.00 \\
        \cmidrule(){1-10}

        \multirow{6}{*}{DeepAE} & Baseline & 9.31 & 9.56 & 9.24 & 9.20 & 9.03 & 3.26 & 15.50 & 44.61 \\ 
        & LR-BA \cite{gu2023lrba} & 15.05 &12.72 &51.74 & 49.62& 46.07& 0& 0& 82.99\\
        & HijackVFL \cite{qiu2024hijack} & 0 & 0& 0&0 &0 &0 & 0& 0\\
        & ADI \cite{pang2023adi} & 13.53 & 16.19 & 74.83& 40.14& 60.58& 0.01& 15.34& 83.31 \\
        & TIFS'23 \cite{he2023backdoor} & 35.32&60.02 &84.05 &17.45 & 80.74& 29.98& 40.25& \textbf{100.0} \\
        & \textbf{VTarbel (Ours)} & \textbf{38.84} & \textbf{89.87} & \textbf{89.05} & \textbf{90.39} & \textbf{90.28} & \textbf{40.77} & \textbf{90.97} & 96.05\\
        
        \hhline{==========}
        \multirow{6}{*}{KDE} & Baseline & 9.25 & 9.60 & 9.27 & 8.60 & 9.10 & 3.26 & 14.62 &44.32 \\ 
        & LR-BA \cite{gu2023lrba} & 17.00& 17.51& 31.99& 43.49& 52.21& 1.69& 2.88& 87.64\\
        & HijackVFL \cite{qiu2024hijack} &0 &6.88 &57.05 & 0&58.14 &69.18 & 22.20& 0\\
        & ADI \cite{pang2023adi} &16.13 &21.69 &76.67 &40.28 &71.56 &4.57 &12.82 & 55.32\\
        & TIFS'23 \cite{he2023backdoor} & \textbf{38.57} &50.16 & 79.98& 0& 78.83& 32.25& 35.74& 92.00 \\
        & \textbf{VTarbel (Ours)} & 38.44 & \textbf{89.54} & \textbf{88.28} & \textbf{90.28} & \textbf{88.21} & \textbf{52.55} & \textbf{90.97} & \textbf{95.84}\\

        \bottomrule
        
    \end{tabular}
    \begin{tablenotes}[flushleft]
        \footnotesize
        \item The ``Ground-Truth'' row denotes the proportion of the number of targeted samples in the test set. The highest ASR values in each column are highlighted in \textbf{bold}.
    \end{tablenotes}
    \end{threeparttable}

    }
    \label{tab:attack_performance}
\end{table}

\subsection{Attack Performance and Comparative Analysis} \label{sec:attack_effectiveness}
In this section, we evaluate the attack performance of VTarbel and compare it with other attack methods. 
As shown in Table ~\ref{tab:attack_performance}, VTarbel consistently achieves superior attack performance in nearly all cases, with the ASR from VTarbel being significantly higher than both the baseline and other attacks.
For example, on the CIFAR10 dataset with the DeepAE detector, VTarbel increases the ASR from the baseline's 9.24\% and 9.03\% (for the VGG13 and ResNet18 models, respectively) to a much higher level of 89.05\% and 90.28\%, respectively. 
Moreover, when compared to other attack methods, VTarbel outperforms them by a significant margin. 
For instance, in the case of (DistilBERT, TREC, KDE), VTarbel achieves an ASR of 90.97\%, whereas the highest ASR from other methods, TIFS'23, is only 35.74\%. 
This demonstrates a substantial boost in attack performance.

For other attack methods, we observe that they struggle to achieve high ASRs under both DeepAE and KDE detectors. 
Specifically, for HijackVFL, the attack performance is poor under the DeepAE detector, with an ASR of 0 across all cases, even lower than the baseline. 
We believe this is because, although the trigger vector added to the attacker's feature embedding has strong predictive capability for the targeted label, it significantly deviates from the normal distribution of the targeted class's feature embedding. 
This causes the anomaly detector to flag all of the attacker's feature embeddings as anomalies, resulting in an ASR of 0.
Among the other attacks, TIFS'23 performs the best. 
For example, under the KDE detector with the ResNet18 model, TIFS'23 increases the ASR from the baseline's 9.10\% on CIFAR10 and 3.26\% on CIFAR100 to 78.83\% and 32.25\%, respectively. 
We hypothesize that this is due to the trigger embedding being located in the highest density region of the targeted feature embedding space, which has a high probability of being mapped to the target label while avoiding detection. 
However, even in these cases, there remains a significant performance gap compared to VTarbel, which achieves much higher ASRs of 88.21\% and 52.55\%, respectively.

These results highlight that, despite requiring minimal attack knowledge compared to existing methods, VTarbel effectively leverages the information available during the VFL inference phase to train both surrogate model and estimated detector.
The surrogate model guides the optimization of malicious feature embeddings, while the estimated detector constrains the embeddings to avoid detection. 
As a result, VTarbel consistently achieves superior performance compared to prior attacks.

\begin{table}
    \footnotesize
    \caption{Comparison of attack performance (ASR (\%)) among VTarbel and its variants.}
    \centering
    \begin{tabular}{c||c|c|c|c}
        \toprule
        \multirow{2}{*}{Method} & MLP3 & VGG13 & ResNet18 & DistilBERT \\
        & TabFMNIST & CIFAR10 & CIFAR10 & TREC \\
        \midrule
        Ground-Truth & 10.00 & 10.00 & 10.00 & 16.90 \\
        Only-Preparation (baseline) & 9.60 & 9.24 & 9.03 & 15.50\\
        Only-Attack & 0 & 0 & 0 & 0\\
        \makecell{Random-Preparation (w/ clustering)} &  84.66 &83.73 &  82.34 & 73.10\\
        \makecell{Random-Preparation (w/o clustering)} & 67.48 & 81.74 & 81.67 & 64.25\\
        Random-Attack & 35.82 &60.26 &40.73 &  73.14\\
        \cmidrule{1-5}
        \textbf{VTarbel} & \textbf{89.87} &\textbf{89.05} & \textbf{90.28} & \textbf{90.97}\\
        \bottomrule

    \end{tabular}
    \label{tab:stage_ablation}
\end{table}

\subsection{Stage Effectiveness Breakdown} \label{sec:stage_ablation}
In this section, we break down and evaluate the effectiveness of each stage in VTarbel, comparing it with five variants, all evaluated under defender's DeepAE detector, as shown in Table~\ref{tab:stage_ablation}. 
These variants are:
(1) Only-Preparation: Consists only of the preparation stage, corresponding to the baseline.
(2) Only-Attack: Involves only the attack stage, where the attacker begins the attack at the very start of VFL inference phase.
(3) Random-Preparation (w/ clustering): The attacker selects the same number of test samples within each cluster as VTarbel, but randomly.
(4) Random-Preparation (w/o clustering): The attacker randomly selects the same total number of test samples as VTarbel from the entire test set, without applying clustering.
(5) Random-Attack: The attacker randomly selects samples predicted as targeted label in the preparation stage as attack samples for the attack stage, rather than optimizing the attack samples.

From Table ~\ref{tab:stage_ablation}, we first observe that attack variants consisting of only one stage, i.e., Only-Preparation and Only-Attack, exhibit extremely low ASR. 
Specifically, Only-Attack consistently results in an ASR of 0 across all cases. 
This occurs because the surrogate model is completely random, meaning the feature embedding optimized from this random model lacks any predictive capability for the targeted label. 
These findings underscore the necessity of our two-stage framework for an effective attack.

Additionally, when we examine attack variants that include two stages but with one stage involving random operations, we find higher ASRs than those observed with only one stage. 
However, there is still a performance gap compared to VTarbel.
For instance, in the (DistilBERT, TREC) case, the ASRs from Random-Preparation (w/ clustering) and Random-Preparation (w/o clustering) are 73.10\% and 64.25\%, respectively. 
In contrast, VTarbel achieves an ASR of 90.97\%, demonstrating significantly better performance.
These results validate the effectiveness of the MMD-based sample selection strategy in VTarbel, which outperforms random selection by targeting expressive samples that contribute to a higher ASR. 
Moreover, we observe that Random-Preparation (w/ clustering) consistently outperforms Random-Preparation (w/o clustering) in all cases. 
This suggests that the balanced preparation set, achieved through semi-supervised clustering in the preparation stage, yields a higher ASR than a randomly selected preparation set.
Lastly, VTarbel outperforms Random-Attack in all cases, indicating that the malicious optimized feature embedding has much stronger predictive power for the targeted label compared to the benign feature embedding of the targeted label. 
This demonstrates that the optimization step during the attack stage is critical for effective attack.

\begin{table}
    \footnotesize
    \caption{Verification of the optimality of the preparation set selected by VTarbel in achieving the highest ASR (\%), compared with variants of early stopping and late stopping during the preparation stage.}
    \centering
    
    \begin{tabular}{cccc||cc|c|cc}
        \toprule
        \multirow{2}{*}{Model} & \multirow{2}{*}{Dataset} & \multirow{2}{*}{\makecell{No. Test \\ Samples}} & \multirow{2}{*}{\makecell{$|\mathcal{Q}^*|$}} & \multicolumn{2}{c|}{Early Stopping} & Best & \multicolumn{2}{c}{Late Stopping} \\
        & & & & -30\% & -15\% & 0 & +15\% & +30\% \\
        \midrule
        MLP3 & TabFMNIST & 10,000 & 900 &82.28 & 87.19 & \textbf{89.87} & 85.41 & 86.76 \\
        VGG13 & CIFAR10 & 10,000 & 1,100 & 88.92 & 88.63 & \textbf{89.05} & 88.78 & 88.77 \\
        ResNet18 & CIFAR10 & 10,000 & 1,000 
        & 77.08 &81.97 & \textbf{90.28} &83.28 &76.37 \\
        DistilBERT & TREC & 554 & 60 & 70.37 & \textbf{93.32} & 90.97 & 85.74 & 81.94\\       
        \bottomrule

    \end{tabular}
    \label{tab:optimality_verification}
\end{table}

\subsection{Optimality Verification of Preparation Set}
As outlined in Section ~\ref{sec:two_stage_formulation}, directly solving the combinatorial problem in Eq.(\ref{eq:combinatorial_problem}) is NP-hard. 
Therefore, our approach selects the minimum number of expressive samples necessary to train an accurate surrogate model and estimated detector, aiming to achieve a high ASR during the attack stage.
In this section, we verify the optimality of the preparation set selected by VTarbel. 
As shown in Table ~\ref{tab:optimality_verification}, we first list the preparation set size \( |\mathcal{Q}^*| \) identified by VTarbel for each evaluation case. 
We then compare the attack performance achieved by VTarbel (denoted as "Best") with two other variants, which differ in the stopping points during the preparation stage: early stopping and late stopping.
For example, the "-15\%" in early stopping means that the preparation stage is terminated when the preparation set size reaches 85\% of the optimal \( |\mathcal{Q}^*| \) found by VTarbel. 
Similarly, the "+15\%" under late stopping means that the preparation stage is extended until the preparation set size reaches 115\% of \( |\mathcal{Q}^*| \). 

We observe that, in most cases, the preparation set selected by VTarbel achieves the optimal ASR. 
For instance, under the (ResNet18, CIFAR10)configuration, VTarbel achieves the highest ASR of 90.28\%, while the early stopping (\(-15\%\)) and late stopping (\(+15\%\)) variants only achieve ASRs of 81.97\% and 83.28\%, respectively, both of which are substantially lower than VTarbel's result.
The reason behind this performance difference is as follows: For late stopping, the optimal \( |\mathcal{Q}^*| \) size provides enough test samples for the surrogate model and estimated detector to be sufficiently trained. 
Adding more samples in the preparation stage reduces the number of samples predicted as the targeted label during the attack stage, thus lowering the ASR. 
In contrast, for early stopping, fewer test samples than VTarbel's optimal preparation set slightly degrade the performance of the estimated detector, leading to a reduced number of samples predicted as targeted label in the attack stage, which also results in a lower ASR compared to VTarbel.


\subsection{Detector-Agnostic Attack}
In this section, we evaluate whether VTarbel is agnostic to the type of detector used by the defender.
Based on the attacker's knowledge of the defender's detector type, we categorize the attack scenarios into white-box and black-box scenarios. 
The former consists of cases where both the active and passive parties use the same detector type, such as (DeepAE, DeepAE) and (KDE, KDE). 
The latter includes scenarios where the attacker's detector type differs from the defender's, such as (DeepAE, KDE) and (KDE, DeepAE), where each pair denotes the (Active Party's Detector, Passive Party's Detector).
From Table ~\ref{tab:detector_agnostic}, we observe that in the black-box attack scenario, where the attacker has no knowledge of the defender's specific detector type, the attacker can still achieve high ASR comparable to those in the white-box attack scenario. 
For example, when the defender's detector is KDE, in the cases of (VGG13, CIFAR10) and (ResNet18, CIFAR10), the ASRs from the black-box attack (where the attacker's detector is DeepAE) are 88.80\% and 90.40\%, respectively, which are even higher than the ASRs of 88.28\% and 88.21\% in the white-box attack scenario.
These results demonstrate that even when the attacker's detector type differs from the defender's, the locally generated malicious feature embedding can still effectively evade detection, highlighting VTarbel's robustness to the specific detector used by the defender.

\begin{table}
    \footnotesize
    \caption{
        VTarbel's attack performance (ASR (\%)) in the white-box and black-box attack scenarios, depending on whether the passive party (attacker) has knowledge of the active party's (defender) detector type.
    }
    \centering
    
    \begin{tabular}{ccc||c|c|c|c}
        \toprule
        \multirow{2}{*}{\makecell{Attack \\Scenario}}  & \multirow{2}{*}{\makecell{Active Party's \\ Detector}} & \multirow{2}{*}{\makecell{Passive Party's \\ Detector}}  & MLP3 & VGG13 & ResNet18 & DistilBERT \\
        & & & TabFMNIST & CIFAR10 & CIFAR10 & TREC \\
        \midrule
        \multirow{2}{*}{\makecell{White-Box}} & DeepAE & DeepAE & 89.87 & 89.05 & 90.28 & 90.97\\
        & KDE & KDE & 89.54 & 88.28 & 88.21 & 90.97\\
        
        \cmidrule{1-7}
        \multirow{2}{*}{\makecell{Black-Box}} & DeepAE & KDE & 89.20 & 88.16 & 88.27 & 90.97\\
        & KDE & DeepAE & 86.93 & 88.80 & 90.40 & 90.97\\       
        \bottomrule
    \end{tabular}
    \label{tab:detector_agnostic}
\end{table}

\begin{figure*}[t!]
	\centering
	\begin{minipage}{0.6\linewidth}
		\centering
		\includegraphics[width=0.9\linewidth, trim=0 460 0 0, clip]{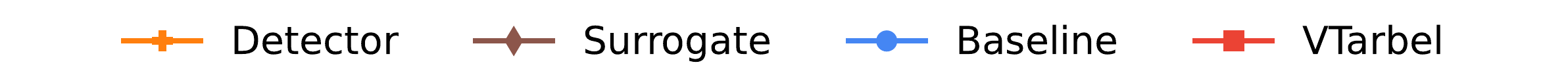}
	\end{minipage}

	\vspace{-2pt}
	
	\begin{minipage}{1.0\linewidth}
		\centering
		\subfigure[TabMNIST]{
        \label{fig:feat_fracs_tab_mnist}
        \includegraphics[width = 0.32\linewidth]{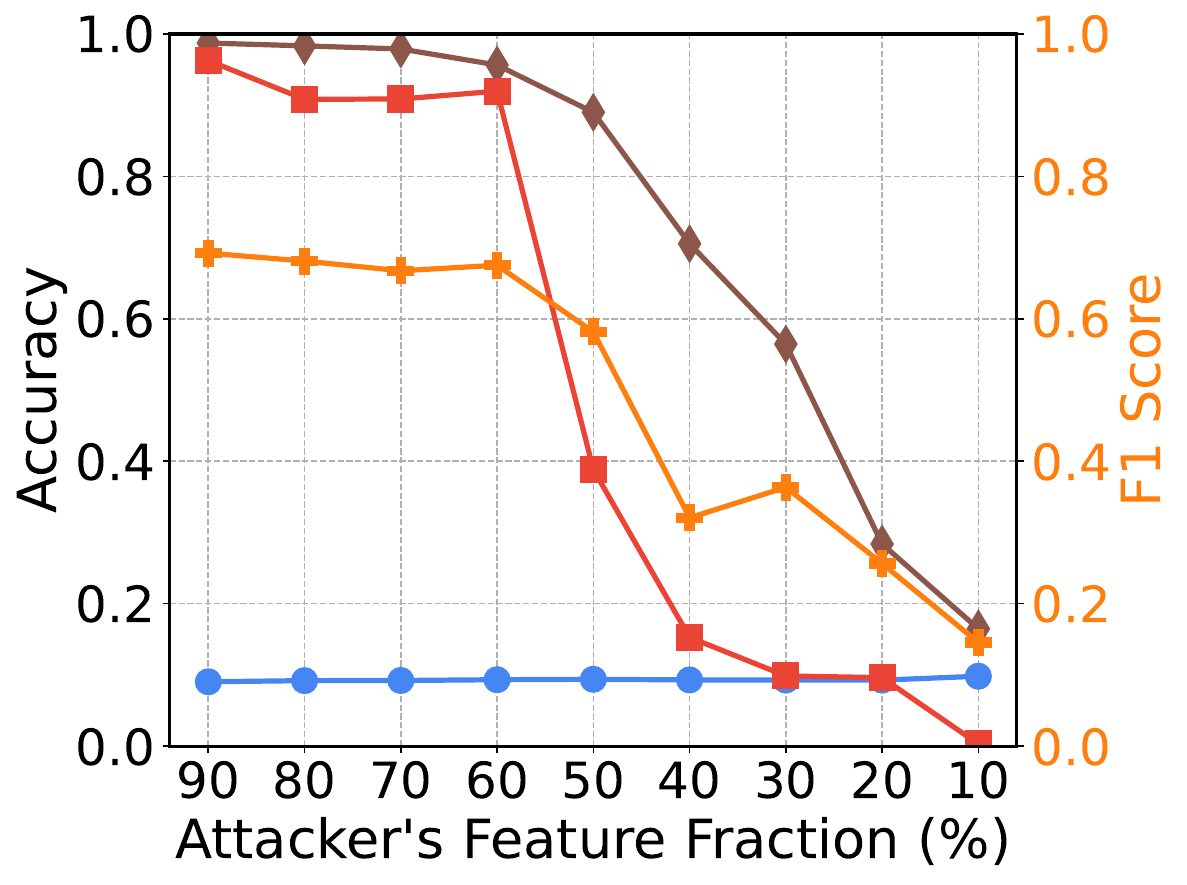}
    }
    \subfigure[TabFMNIST]{
        \label{fig:feat_fracs_tab_fashion_mnist}
        \includegraphics[width = 0.32\linewidth]{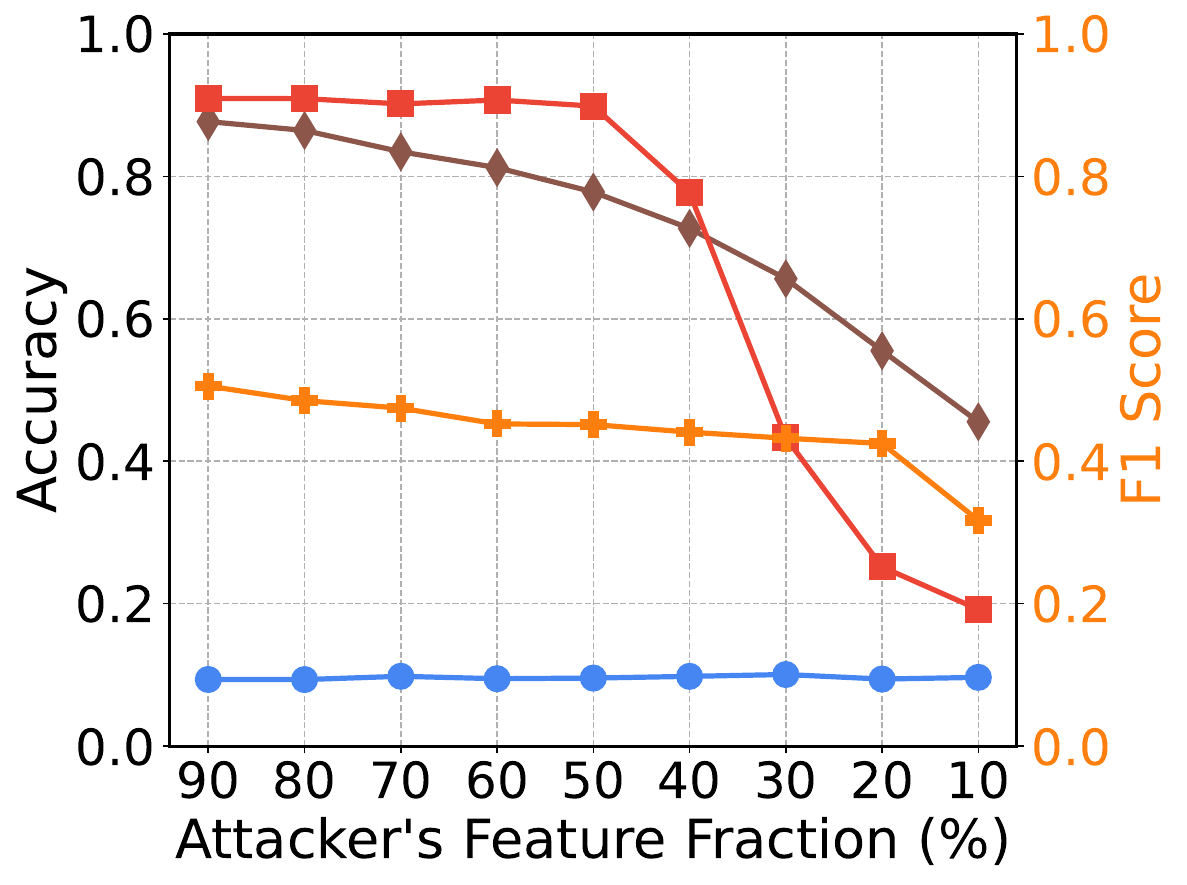}
    }\hfill
		
	\end{minipage}

    \caption{The impact of the attacker's feature fraction in a two-participant VFL inference system, where the features allocated to each participant are randomly selected.}
    \label{fig:feat_fracs}
\end{figure*}

\subsection{Scalability Evaluation}
\begin{figure*}[t!]
	\centering

    \begin{minipage}{0.6\linewidth}
		\centering
		\includegraphics[width=0.9\linewidth, trim=0 460 0 0, clip]{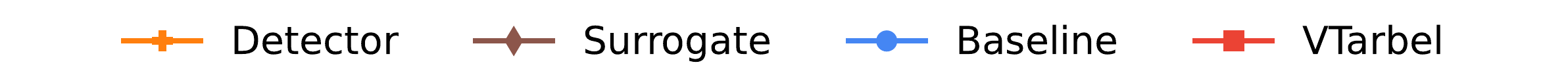}
	\end{minipage}

	\vspace{-2pt}

    \begin{minipage}{1.0\linewidth}
        \centering
        \subfigure[TabMNIST]{
            \label{fig:num_participants_tab_mnist}
            \includegraphics[width = 0.32\linewidth]{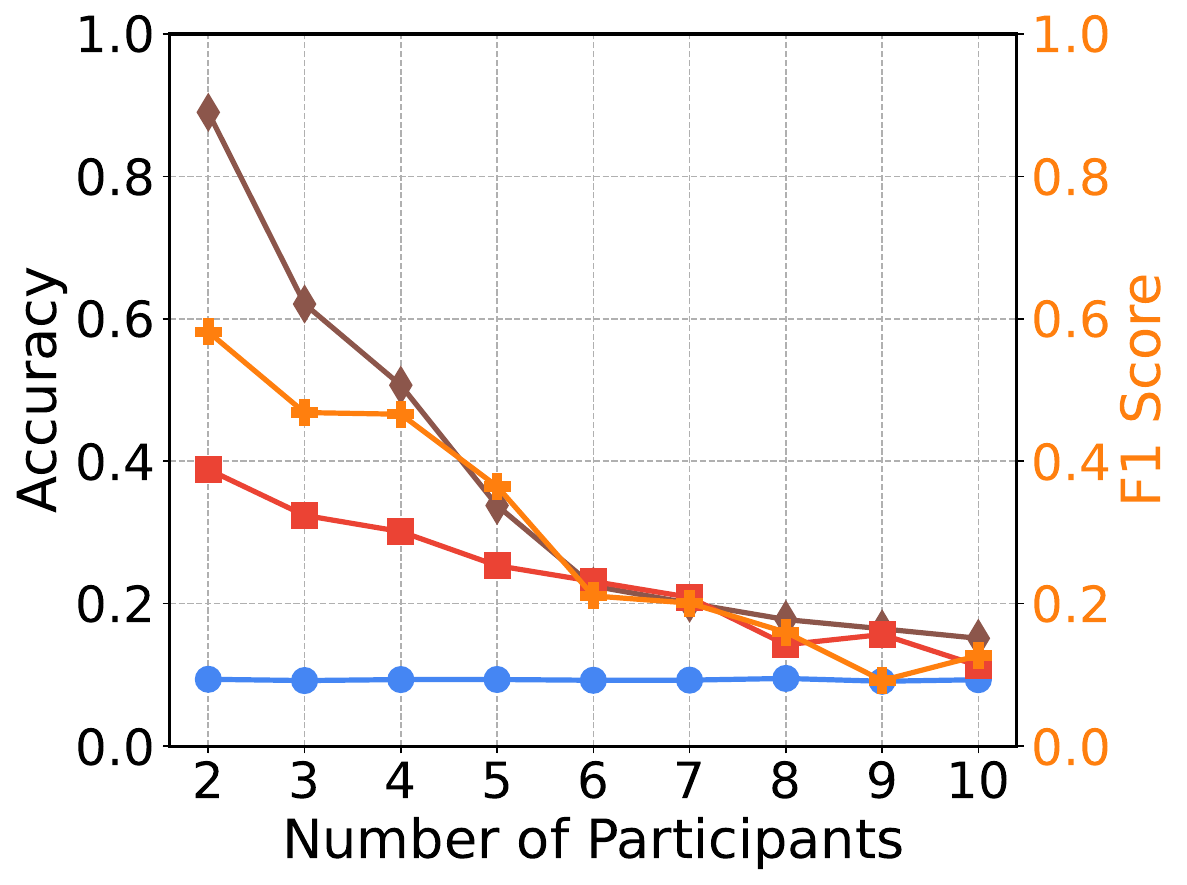}
        }
        \subfigure[TabFMNIST]{
            \label{fig:num_participants_tab_fashion_mnist}
            \includegraphics[width = 0.32\linewidth]{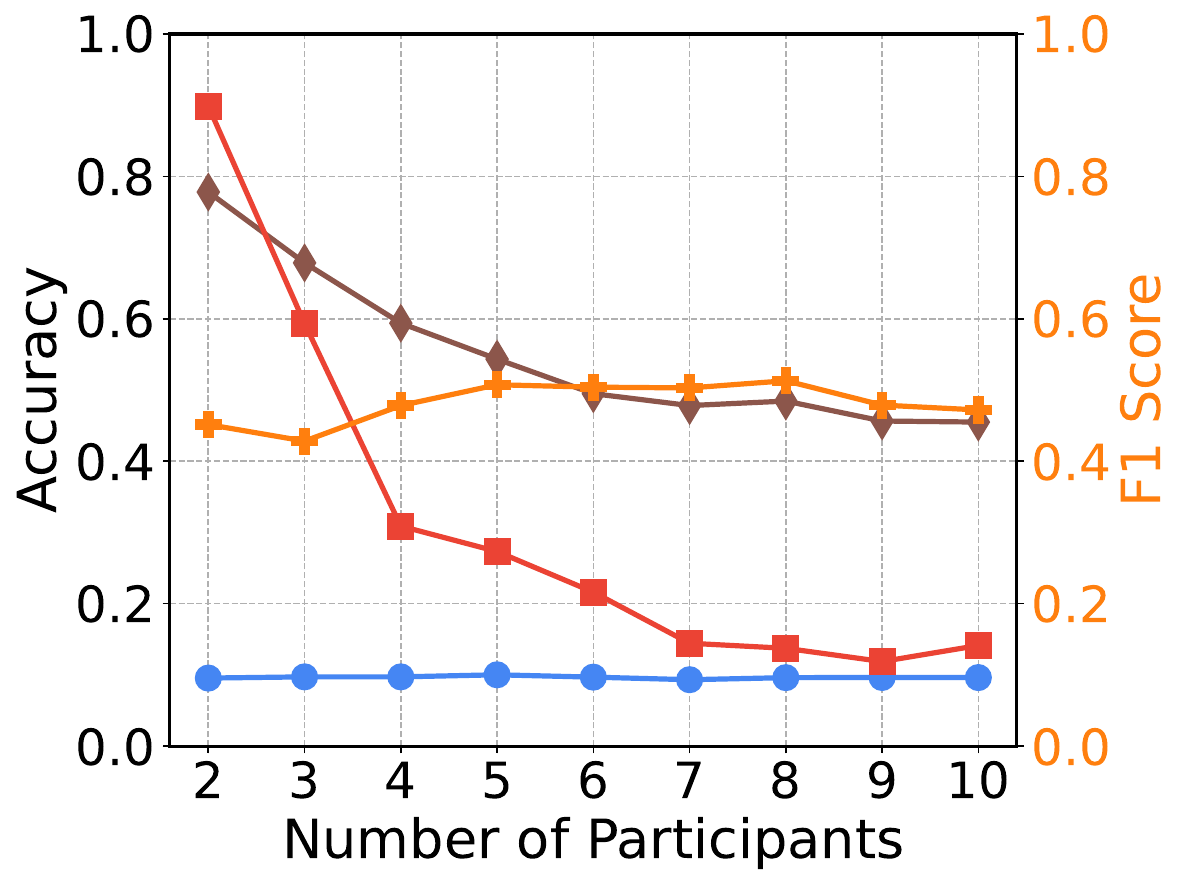}
        }\hfill
    \end{minipage}
	
    \caption{The impact of the number of participants, where the total features are randomly and evenly split among participants.}
    \label{fig:num_participants}
\end{figure*}

In this section, we evaluate VTarbel's attack scalability in terms of the attacker's feature fraction and the total number of participants in the VFL inference system. 
Our experiments are conducted using the TabMNIST and TabFMNIST datasets under two scenarios: the first scenario corresponds to a VFL inference system with two participants, where the attacker's feature fraction (i.e., the fraction of the attacker's features to the total number of features) ranges from 90\% to 10\% (Figure~\ref{fig:feat_fracs}).
The second scenario corresponds to a VFL inference system with varying numbers of participants, ranging from 2 to 10, where the total features are randomly and evenly allocated to each participant (Figure~\ref{fig:num_participants}).

From both Figure \ref{fig:feat_fracs} and Figure \ref{fig:num_participants}, several similar trends can be observed. 
First, the baseline ASR remains stable and is unaffected by the attacker's feature fraction or the number of participants. 
This is because the baseline is only related to the global VFL model's accuracy. 
Next, as the number of features allocated to the attacker decreases (i.e., the decrease in the attacker's feature fraction in Figure~\ref{fig:feat_fracs} and the increase in the total number of participants in Figure~\ref{fig:num_participants}),
both the accuracy of the locally trained surrogate model and the F1 score of the estimated detector gradually decrease. 
The reason for this is that with fewer features, it becomes much more challenging for the attacker to leverage these limited features to train an accurate surrogate model or detector that can precisely estimate the global VFL model or the defender's detector.
Consequently, there is a corresponding gradual decrease in VTarbel's ASR, which is influenced not only by the less accurate surrogate model and detector but also by the benign features from other participants that can neutralize the malicious properties of the attacker's feature embedding.
However, we note that practical VFL systems typically involve only two participants, where the passive party (i.e., the attacker) serves as the feature provider and often possesses more features than the active party (i.e., the defender) \cite{chen2021homomorphic}.
In such practical settings, VTarbel demonstrates strong attack performance. 
For instance, as shown in Figure ~\ref{fig:feat_fracs}, 
when the attacker holds 60\% of the features, the ASR reaches 91.97\% on TabMNIST and 90.73\% on TabFMNIST, 
posing a significant threat to the security of practical VFL inference systems.

\begin{figure*}[t!]
	\centering

    \begin{minipage}{0.8\linewidth}
		\centering
		\includegraphics[width=0.9\linewidth, trim=0 440 0 0, clip]{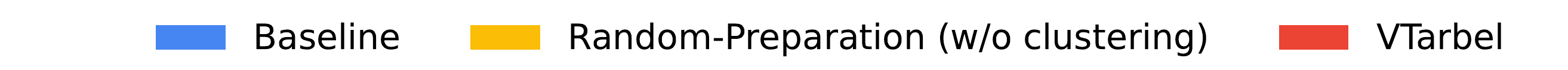}
	\end{minipage}

	\vspace{-2pt}

    \begin{minipage}{1.0\linewidth}
        \centering
        \subfigure[VGG13]{
            \label{fig:ablation_unbalance_dataset_vgg13}
            \includegraphics[width = 0.32\linewidth]{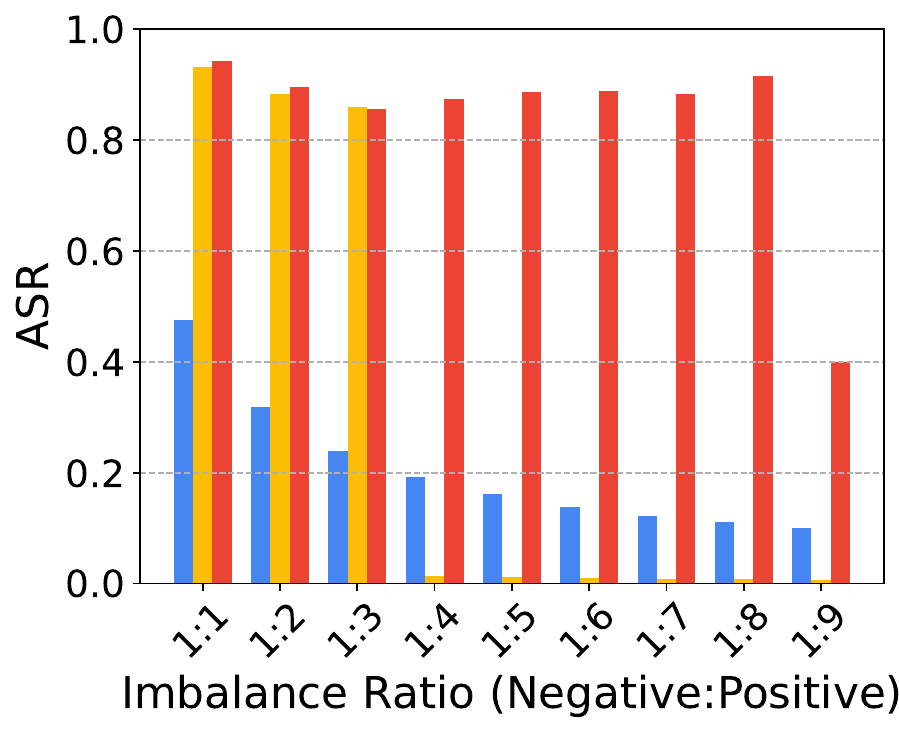}
        }
        \subfigure[ResNet18]{
            \label{fig:ablation_unbalance_dataset_resnet18}
            \includegraphics[width = 0.32\linewidth]{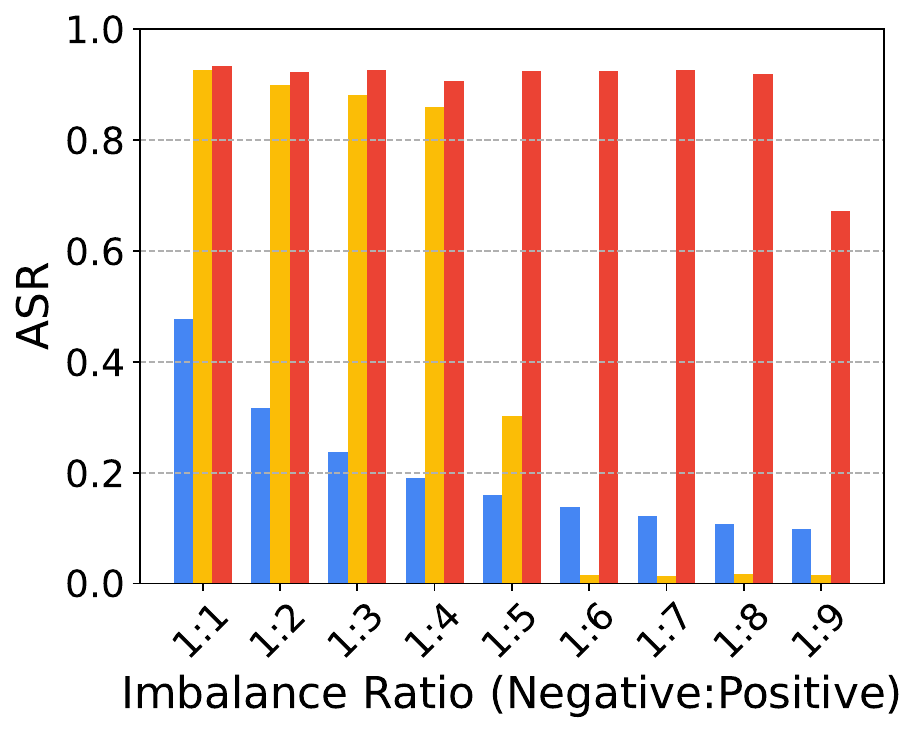}
        }\hfill
    \end{minipage}
	
    \caption{Impact of an imbalance binary dataset, where the negative class is the targeted label.}
    \label{fig:ablation_unbalance_dataset}
\end{figure*}

\subsection{Ablation Study} \label{sec:ablation_study}

\subsubsection{Imbalanced Dataset}
We investigate the impact of an imbalanced dataset on VTarbel's attack performance. 
We construct a binary test set containing 1,000 samples by randomly selecting test samples from class 0 (treated as the negative class) and class 1 (treated as the positive class) from the CIFAR10 dataset, and evaluate it using the VGG13 and ResNet18 models. 
As shown in Figure~\ref{fig:ablation_unbalance_dataset}, the imbalance ratio on the x-axis is defined as the number of negative samples divided by the number of positive samples, where the targeted label is the negative class. 
For VTarbel, we observe that even as the imbalance ratio increases from 1:1 to 1:8, the ASR remains stable and comparable to that of the balanced dataset.
For example, when the imbalance ratio is 1:8, the ASR for VGG13 and ResNet18 is 0.916 and 0.919, respectively, both of which are close to the ASRs of 0.942 and 0.934 observed when the imbalance ratio is 1:1.
This demonstrates that VTarbel is robust to class imbalance within the test set.
Furthermore, when compared to the variant Random-Preparation (w/o clustering), we find that when the class imbalance becomes more severe (e.g., the imbalance ratio is greater than 1:5), the ASR for Random-Preparation (w/o clustering) is close to zero and even lower than that of the baseline. 
In contrast, VTarbel consistently achieves a considerably higher ASR than Random-Preparation (w/o clustering). 
This further validates that the clustering step in the preparation stage is crucial, as it ensures that the preparation set remains balanced, which facilitates the training of both the surrogate model and the estimated detector, even when the original test set is imbalanced.

\begin{figure*}[t!]
    \centering

    \begin{minipage}{1.0\linewidth}
		\centering
		\includegraphics[width=0.95\linewidth, trim=0 450 0 0, clip]{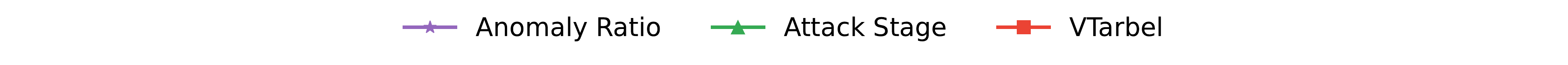}
	\end{minipage}

	\vspace{-2pt}

    \begin{minipage}{1.0\linewidth}
        \centering
        \subfigure[Varying defender's thresholds]{
		\label{fig:active_percentile}
		\includegraphics[width = 0.32\textwidth]{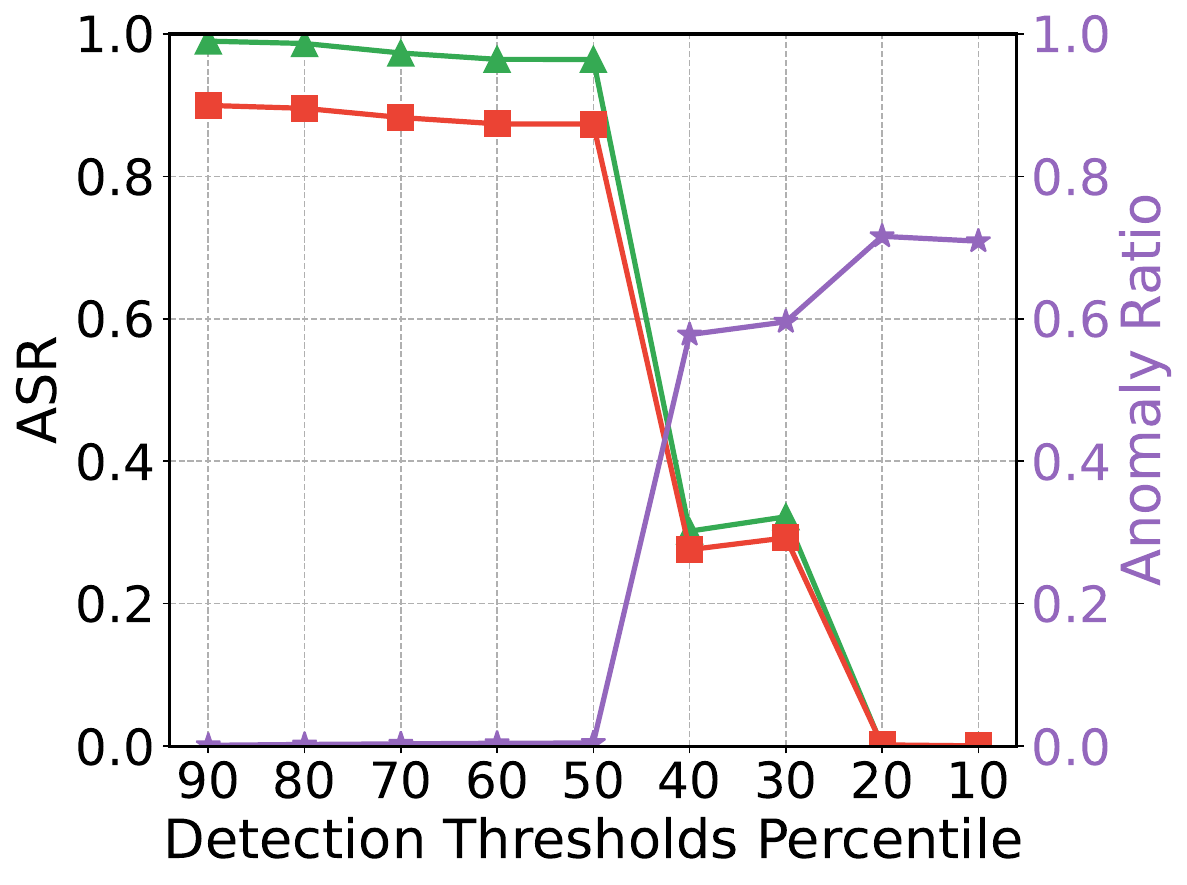}
        }\hfill
        \subfigure[Varying attacker's thresholds]{
            \label{fig:passive_percentile}
            \includegraphics[width = 0.32\textwidth]{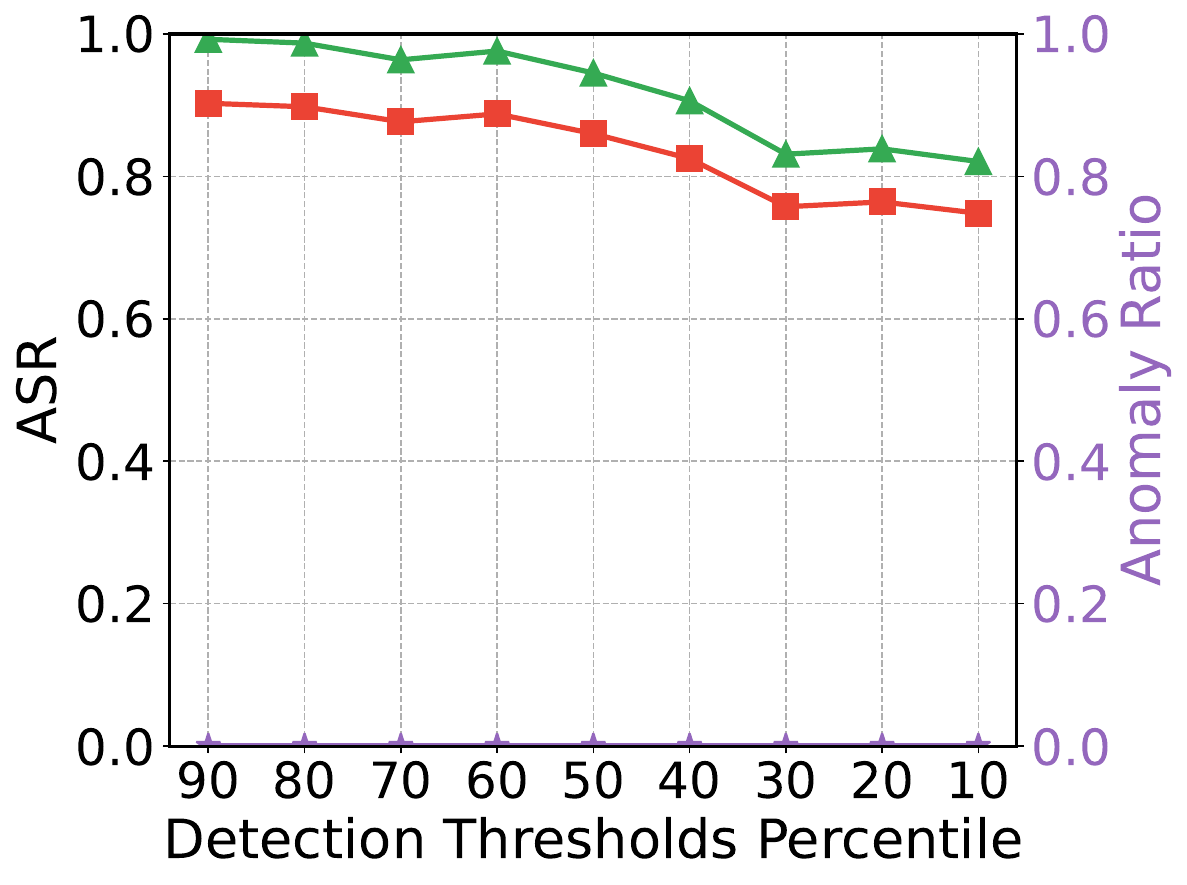}
        }\hfill
        \subfigure[Varying both thresholds]{
            \label{fig:both_percentile}
            \includegraphics[width = 0.32\textwidth]{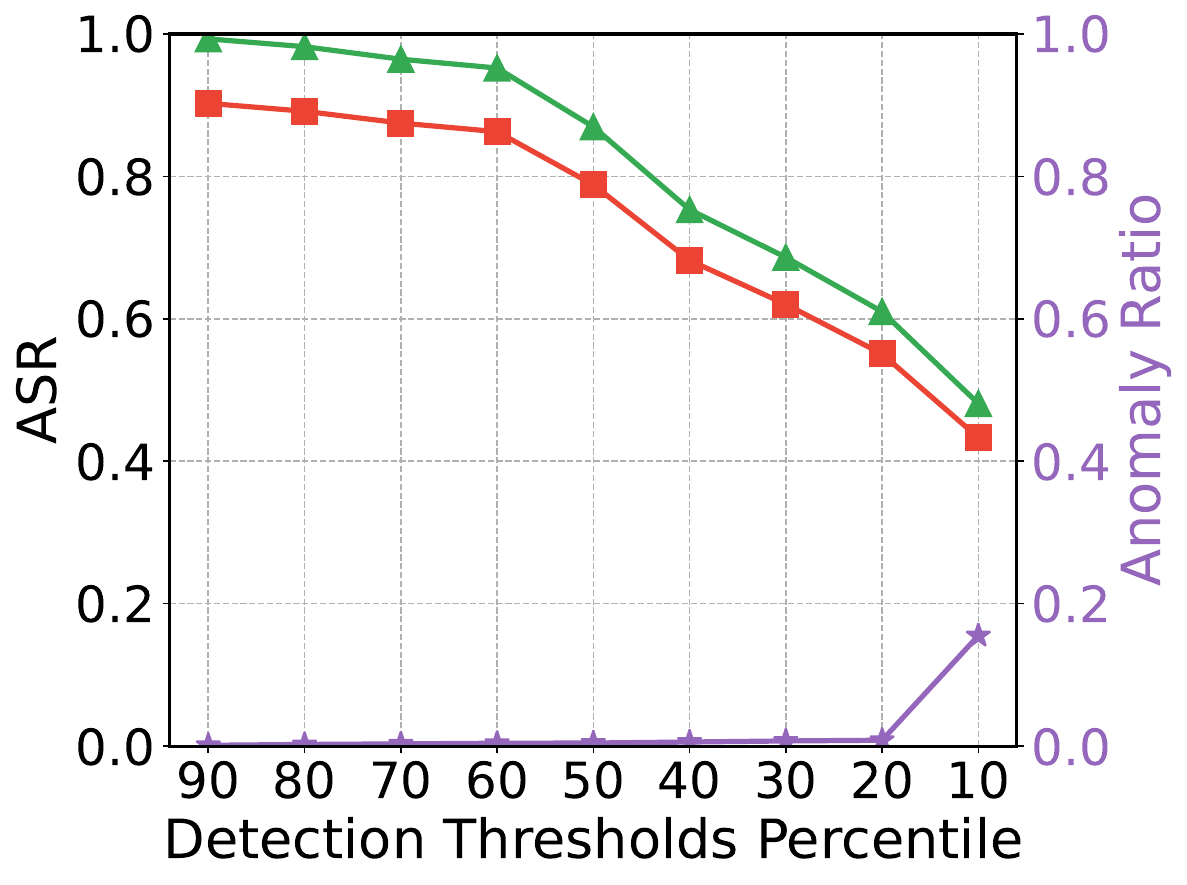}
        }\hfill
    \end{minipage}

	\caption{
		Impact of anomaly detection threshold under the (ResNet18, CIFAR10, DeepAE) configuration.
	}
	\label{fig:detection_threshold}
\end{figure*}

\subsubsection{Anomaly Detection Threshold}
We evaluate the impact of anomaly detection thresholds on VTarbel's attack performance. 
Specifically, we investigate three different cases based on the percentiles of the detection thresholds of both the defender's deployed detector and the attacker's estimated detector:
(1) The defender's percentile of detection thresholds  ranges from 90 to 10, implying gradually strengthened detection, while the attacker's percentile of detection threshold is fixed at the default 95 (Figure~\ref{fig:active_percentile}).
(2) The attacker's percentile of detection thresholds  ranges from 90 to 10, indicating more conservative optimization of the feature embedding. 
In this case, the optimization process may be early terminated due to the strengthened estimated detection threshold, while the defender's detection threshold percentile is fixed at the default 95 (Figure~\ref{fig:passive_percentile}).
(3) Both the defender's and attacker's percentiles of detection threshold range from 90 to 10 simultaneously  (Figure~\ref{fig:both_percentile}).
Note that the defender's detection thresholds percentiles are calculated based on the training set, while the attacker's are calculated based on the preparation set. 
The experiments are conducted using the ResNet18 model and the CIFAR10 dataset under the DeepAE detector.

As shown in Figure~\ref{fig:detection_threshold}, we plot the overall ASR of VTarbel, along with the ASR of the attack stage (since the detection thresholds have no effect on the ASR of the preparation stage) and the overall anomaly ratio identified by the defender's detector.
In Figure~\ref{fig:active_percentile}, as we gradually increase the strength of the defender's anomaly detector (e.g., with detection threshold percentiles lower than 50), there is a gradual increase in the anomaly ratio. 
This occurs because most of the attacker's optimized feature embeddings are recognized as anomalies. 
As a result, the ASR of the attack stage gradually decreases, and the overall ASR of VTarbel also decreases simultaneously.
In contrast, Figure~\ref{fig:passive_percentile} shows that as the detection strength of the attacker's estimated detector increases, the anomaly ratio remains stable and close to 0. 
The reason for this is that as the detection strength of the estimated detector increases, the optimization step in Algorithm~\ref{algo:attack} is terminated earlier with fewer optimization rounds, ensuring that the generated feature embedding stays within a region with high probability. 
Thus, the anomaly ratios consistently remain close to zero.
At the same time, due to fewer optimization rounds, the malicious feature embeddings are less effective at being mapped to the targeted label, which explains the slight decrease in the ASR of the attack stage and the overall ASR of VTarbel.

Finally, in Figure~\ref{fig:both_percentile}, when both the defender's and the attacker's detection threshold percentiles change simultaneously, we observe that the overall ASR of VTarbel lies between the values shown in Figure~\ref{fig:active_percentile} and Figure~\ref{fig:passive_percentile}. 
For instance, when the detection threshold percentiles equal 30, VTarbel's ASR in Figure~\ref{fig:both_percentile} is 62.03\%, which lies between 29.25\% in Figure~\ref{fig:active_percentile} and 75.74\% in Figure~\ref{fig:passive_percentile}. 
This is because, compared with Figure~\ref{fig:active_percentile}, the optimization rounds in Figure~\ref{fig:both_percentile} are fewer, which ensures that most feature embeddings are not recognized as anomalies, thus the overall ASR is higher. 
Furthermore, compared with Figure~\ref{fig:passive_percentile}, the defender's detection strength in Figure~\ref{fig:both_percentile} is stronger, which results in a higher anomaly ratio (0.72\% compared to 0.05\%), leading to a slightly lower overall ASR.


\subsubsection{Distance Constraint}
\begin{figure}[t!]
	\begin{minipage}[t]{0.32\linewidth}
		\centering
		\includegraphics[width=\linewidth, clip]{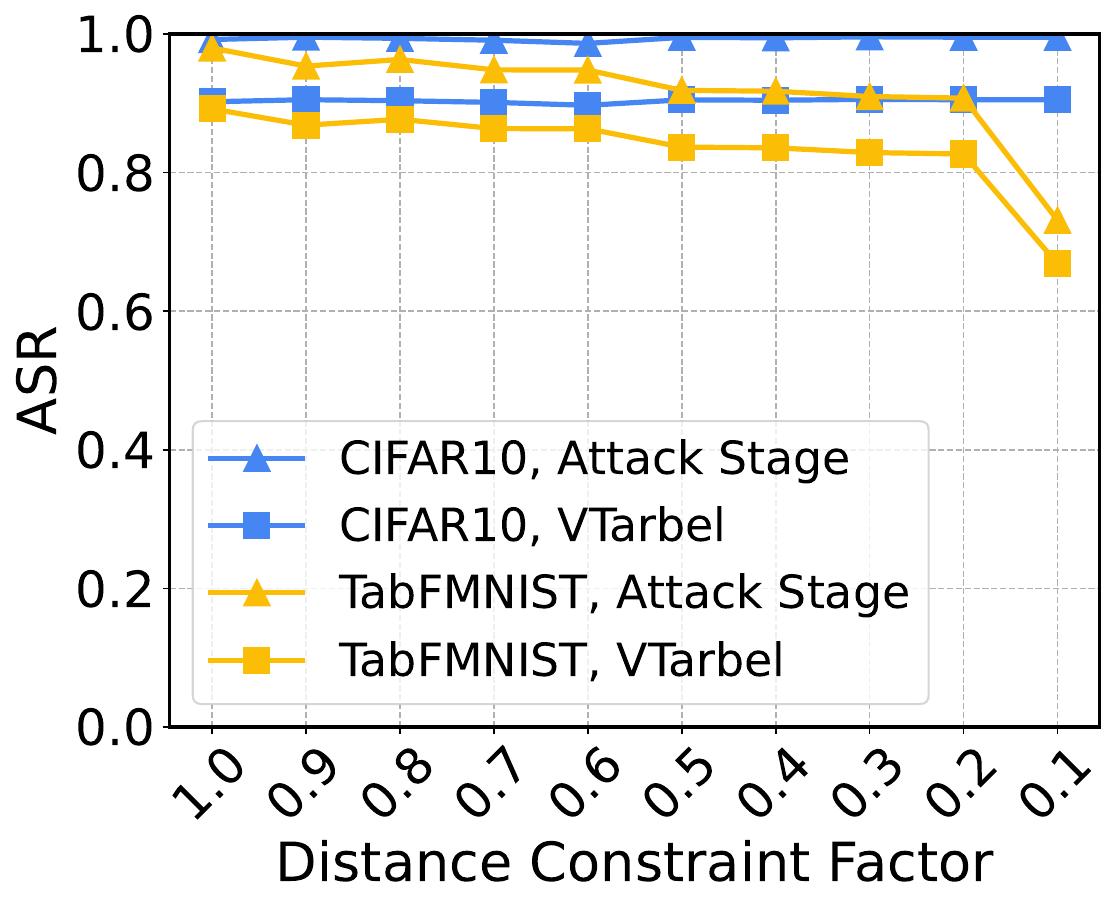}
		\caption{
            Impact of distance constraint.
        }
		\label{fig:ablation_distance_constraint}
	\end{minipage}
    \hspace{15pt}
	\begin{minipage}[t]{0.32\linewidth}
		\centering
		\includegraphics[width=\linewidth, clip]{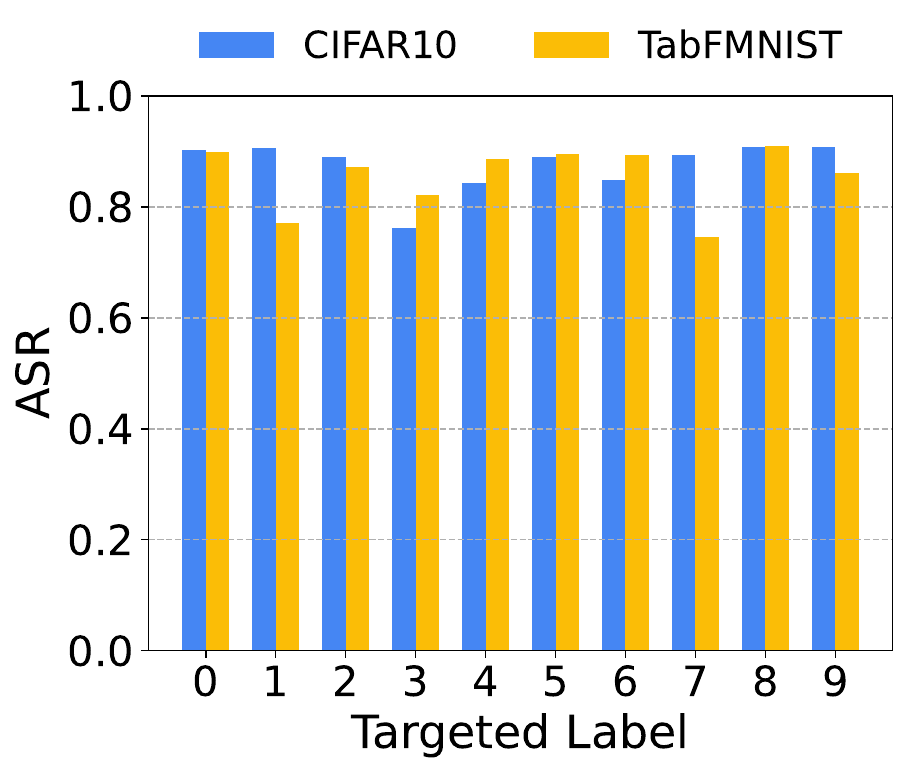}
		\caption{
            Impact of targeted label.
        }
		\label{fig:ablation_targeted_label}
	\end{minipage}
    
    \hfill
\end{figure}

We examine the impact of the distance constraint \( r_{max} \) in the attack stage on VTarbel's attack performance. 
The \( r_{max} \) is computed as the \( L_2 \) distance between the maximum and minimum values in each dimension of the feature embedding of test samples in the preparation set.
As shown in Figure~\ref{fig:ablation_distance_constraint}, on the x-axis, we introduce a \emph{distance constraint factor} \( \beta \in (0, 1] \), which scales the maximum allowed perturbation radius \( r_{max} \). 
The actual distance bound for optimization is then defined as \( \beta r_{max} \).
We observe that as \( \beta \) gradually decreases, for the CIFAR10 dataset (using the ResNet18 model), there is little impact on both the ASR of the attack stage and the overall ASR of VTarbel. 
This demonstrates that VTarbel is highly robust to the allowed perturbation radius.
For the TabFMNIST dataset, there is a gradual decrease in the ASR of the attack stage as \( \beta \) decreases. 
This is because the strengthened perturbation distance constraint restricts the malicious feature embeddings, preventing them from deviating too far from the normal feature space. 
As a result, it becomes less effective at changing the VFL prediction to the targeted label.
Nevertheless, we note that even when \( \beta \) is lowered to 0.1, for the TabFMNIST dataset, the overall ASR of VTarbel remains high at 66.82\%, still posing a considerable threat to the security of the VFL inference system.

\subsubsection{Targeted Label}
We evaluate the impact of the targeted label on VTarbel's attack performance. 
As shown in Figure \ref{fig:ablation_targeted_label}, for both the CIFAR10 and TabFMNIST datasets, VTarbel consistently achieves high ASR across all targeted labels ranging from 0 to 9.
For the CIFAR10 dataset, the minimum, maximum, and average ASRs across all targeted labels are 76.19\%, 90.89\%, and 87.54\%, respectively. 
In the TabFMNIST dataset, the corresponding ASRs are 74.66\%, 90.92\%, and 85.57\%, respectively.
These results confirm that VTarbel is general and agnostic to the specific targeted label.

\begin{figure*}[t!]
	\centering

    \begin{minipage}{1.0\linewidth}
		\centering
		\includegraphics[width=0.95\linewidth, trim=0 490 0 0, clip]{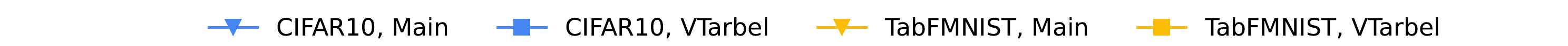}
	\end{minipage}

	\vspace{-2pt}

    \begin{minipage}{1.0\linewidth}
        \centering
        \subfigure[Noisy Embedding]{
            \label{fig:defense_ng}
            \includegraphics[width = 0.32\textwidth]{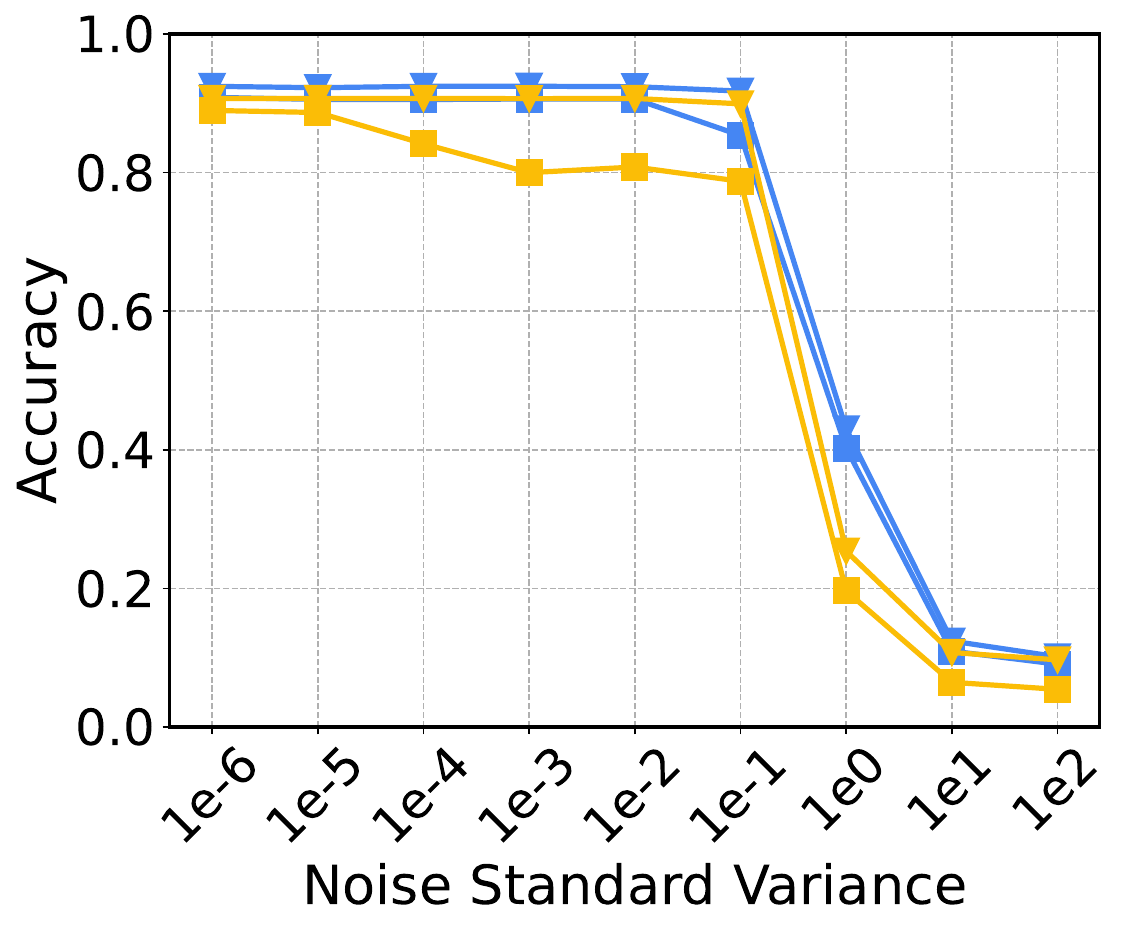}
        }\hfill
        \subfigure[Discrete Embedding]{
            \label{fig:defense_dg}
            \includegraphics[width = 0.32\textwidth]{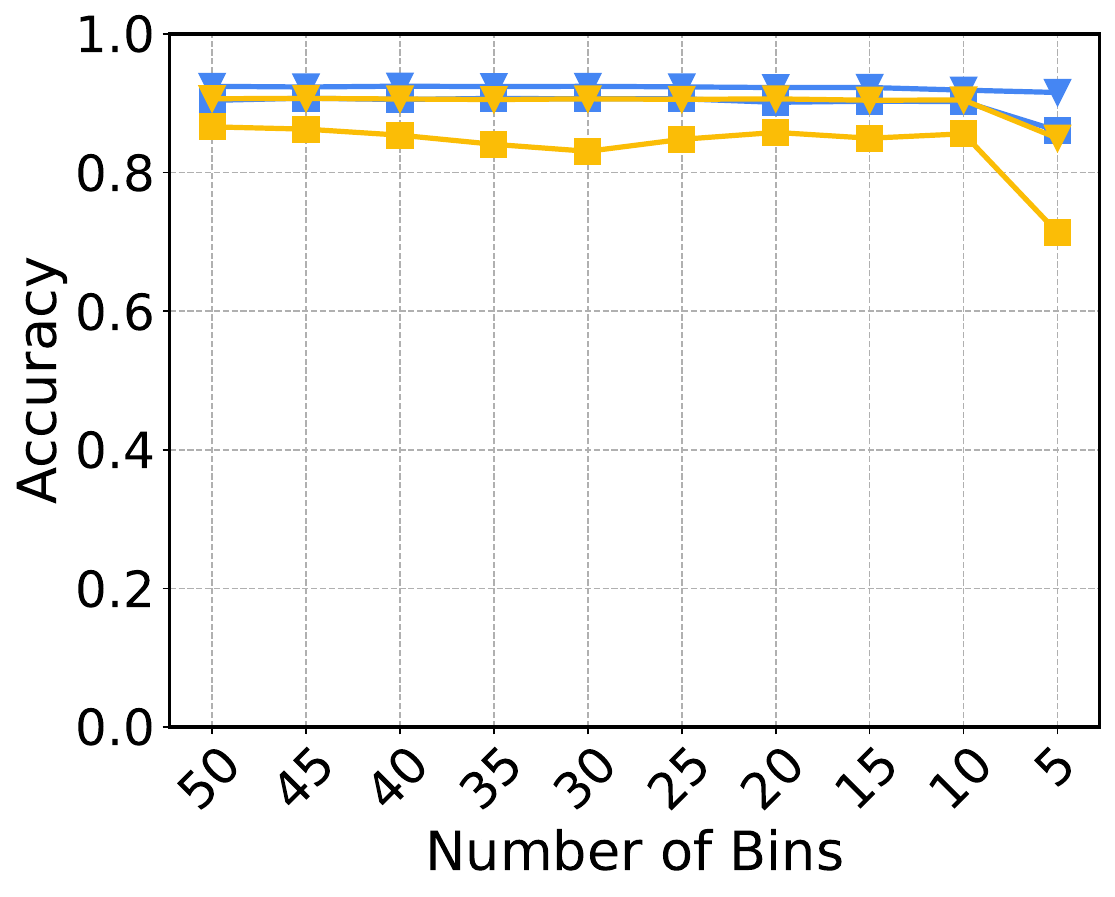}
        }\hfill
        \subfigure[Compressed Embedding]{
            \label{fig:defense_cg}
            \includegraphics[width = 0.32\textwidth]{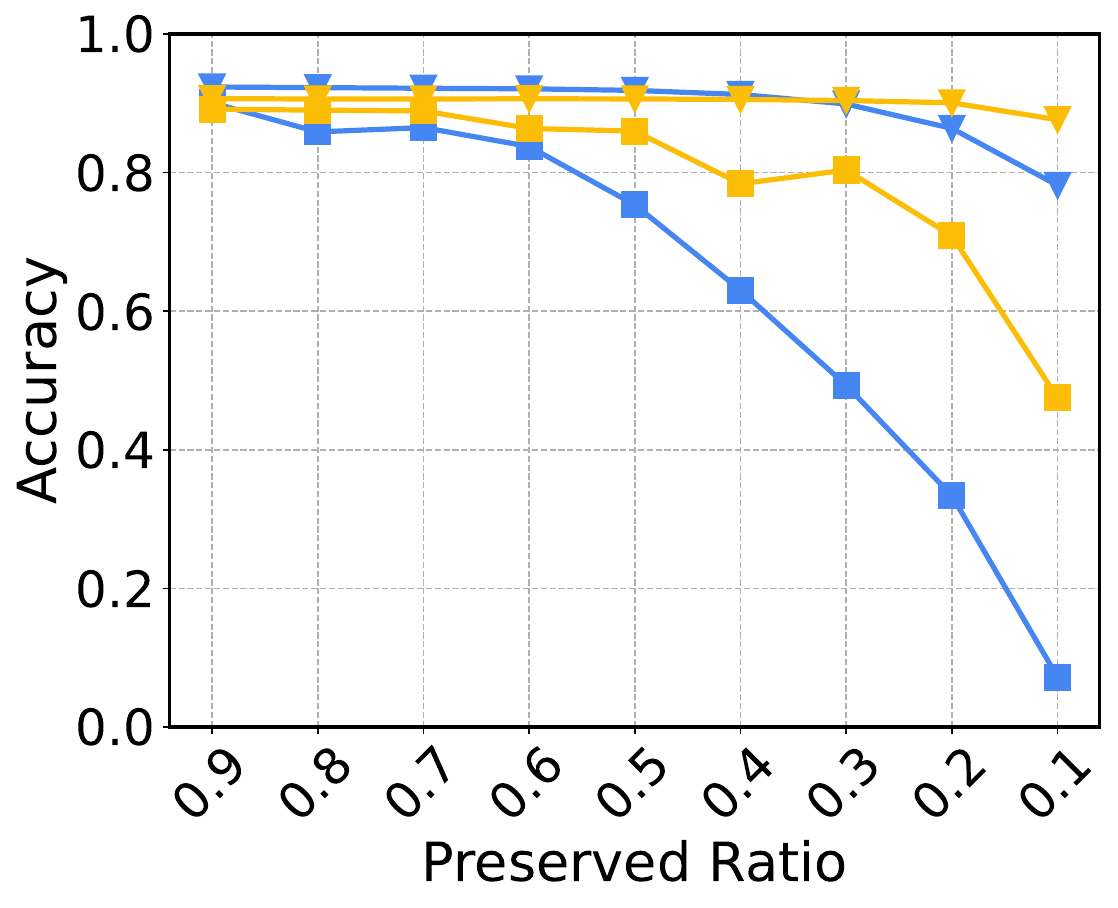}
        }\hfill
    \end{minipage}

	\caption{
		Evaluation of three commonly used privacy-preserving techniques on defending against VTarbel while maintaining main task accuracy.
	}
	\label{fig:defense}
\end{figure*}

\vspace{-4pt}
\subsection{Possible Defenses}
To defend against VTarbel, we validate whether the defender can apply three commonly used privacy-preserving techniques \cite{fu2022label,zhu2019deep} in the ML community to mitigate the targeted label attack while maintaining the main task (i.e., VFL inference) accuracy. 
The experiments are conducted using the (ResNet18, CIFAR10) and (MLP3, TabFMNIST) configurations, all under the DeepAE detector. 
Each defense technique and its corresponding results are discussed below.

\textbf{Noisy Embedding.}
In this defense, the defender adds random Gaussian noise to the received feature embeddings from the attacker. 
The Gaussian noise mean is fixed at zero, and the standard variance ranges from \(1e^{-6}\) to \(1e^2\), as shown in Figure~\ref{fig:defense_ng}. 
We observe that as the noise standard variance increases, VTarbel's ASR gradually decreases in both the CIFAR10 and TabFMNIST datasets. 
However, there is also a simultaneous decrease in main task accuracy, indicating that noisy embedding is not an effective defense strategy.

\textbf{Discrete Embedding.}
In this technique, the defender sets an interval between the minimum and maximum value of the attacker's feature embeddings from training samples, and then slices this interval into several bins. 
At the VFL inference phase, upon receiving feature embeddings from the attacker, the defender rounds each embedding value to the nearest endpoint of the corresponding bin. 
The number of bins controls how much information within the feature embedding is preserved, with a larger number of bins corresponding to less preserved information.
As illustrated in Figure~\ref{fig:defense_dg}, as the number of bins decreases, there is no significant decline in VTarbel's ASR for both the CIFAR10 and TabFMNIST datasets. 
For example, for CIFAR10 dataset, when there are 50 bins, the ASR is 92.45\%, and when there are only 5 bins, the ASR is still high at 91.56\%. 
This indicates that the discrete embedding technique is also not effective in reducing ASR.

\textbf{Compressed Embedding.}
In this technique, for each attacker's feature embedding vector, the defender preserves only a proportion of the largest absolute values, setting all other values to zero. 
As shown in Figure~\ref{fig:defense_cg}, for instance, when the preserved ratio is 0.2, it means that the top 20\% of the largest absolute values of the embedding vector are retained, while the remaining 80\% are set to zero.
We find that the compressed embedding technique outperforms both the noisy embedding and discrete embedding techniques. 
For example, when the preserved ratio decreases from 0.9 to 0.1 in the TabFMNIST dataset, VTarbel's ASR decreases from 89.15\% to 47.50\%, while the main task accuracy only slightly decreases from 90.67\% to 87.64\%. 
Similarly, in the CIFAR10 dataset, VTarbel's ASR decreases from 90.07\% to 7.19\%, while the main task accuracy decreases from 92.35\% to 78.16\%.

From these evaluations,  we conclude that, among these three commonly used techniques, the compressed embedding technique performs the best.
It can, to some extent, preserve main task accuracy while alleviating the attack threat. 
However, a more effective defense strategy that can maintain VFL inference accuracy as in a standard VFL inference system, while mitigating attack accuracy to baseline levels, remains unexplored, which serves as an important direction for future research.

\label{sec:my_exps}


\section{Related Work}


The existing literature highlights various attacks against VFL, which can be broadly categorized into \emph{data inference attacks} and \emph{targeted label attacks}. 
The former, detailed in Section \ref{sec:data_inference_attack}, aim to infer the private data of participants within the VFL system. 
In contrast, the latter, discussed in Section \ref{sec:targeted_output_attack}, focus on manipulating the VFL model to produce attacker-specified label through malicious actions.

\subsection{Data Inference Attacks} \label{sec:data_inference_attack}
Data inference attacks occur when a semi-honest (honest-but-curious) participant exploits the intermediate information exchanged during VFL training to deduce private data belonging to other participants.
These attacks can be further divided into two types: feature reconstruction attacks \cite{he2019model, luo2021feature, weng2020privacy, jin2021cafe, gao2023pcat, pasquini2021unleashing, fu2023focusing} and label inference attacks \cite{fu2022label, li2022label, sun2022label, zou2022label, tan2022residue, kang2022framework}, based on the type of data being inferred.


\textbf{Feature reconstruction attack}
refers to the active party attempting to reconstruct the passive party's raw features from their feature embeddings. 
He et al. \cite{he2019model} develop various approaches to achieve this. 
In a white-box setting, they minimize the distance between actual feature embeddings and dummy embeddings to reconstruct the original raw features. 
Similarly, in a black-box setting, they leverage an inverse network to map feature embeddings back to their corresponding raw features.
Luo et al. \cite{luo2021feature} propose that if both the global VFL model and its predictions are accessible to the attacker, they could design a generator model to minimize the loss between the predictions based on generated features and the actual predictions, effectively transforming the feature reconstruction problem into a feature generation problem.
Pasquini et al. \cite{pasquini2021unleashing} demonstrate that an untrustworthy participant could manipulate the model's learning procedure, leading it to a vulnerable state that facilitates the extraction of other participants' raw features.


\textbf{Label inference attack} 
is typically carried out by the passive party, who aims to infer the active party's private labels by analyzing the backpropagated gradients or their local bottom model. 
Li et al. \cite{li2022label} identify differences in gradient distribution and direction between positive and negative samples in imbalanced binary classification problems. 
Based on these observations, they developed two effective label inference methods. 
Fu et al. \cite{fu2022label} introduce a Model Completion (MC) attack, where the attacker appends an inference head to their well-trained bottom model to create an attack model. 
This model is then fine-tuned with some labeled samples to achieve high accuracy, enabling the inference of labels in both training and testing datasets. 
Furthermore, Tan et al. \cite{tan2022residue} show that an attacker could achieve 100\% accurate label inference by solving a linear system using the local dataset and the received gradients.

\subsection{Targeted Label Attacks} \label{sec:targeted_output_attack}
The targeted label attack involves a malicious participant (usually the passive party) manipulating the VFL model to produce a targeted label of their choosing. 
These attacks can be categorized based on the stage at which they are executed into backdoor attacks \cite{liu2020backdoor, bai2023villain, naseri2024badvfl} and adversarial attacks \cite{qiu2024hijack, gu2023lrba, pang2023adi, he2023backdoor, liu2022copur}.

In \textbf{backdoor attack,}
the attacker stealthily implants a backdoor in the VFL model during training. 
This backdoor is later triggered by specific samples at the inference phase, causing the model to misclassify them as the desired targeted label. 
Typically, attackers begin by conducting a label inference attack (as detailed in Section \ref{sec:data_inference_attack}) to obtain labels, which they use to map trigger samples with the target label. 
Liu et al. \cite{liu2020backdoor} infer labels from intermediate gradients and swap gradients between trigger and target samples to inject the backdoor. 
The authors in \cite{bai2023villain} propose a novel embedding swapping algorithm to deduce labels, subsequently launching a backdoor attack with a stealthy additive trigger embedded within the feature embedding. 
Similarly, Naseri et al. \cite{naseri2024badvfl} introduce BadVFL, which employs the MC attack to infer labels of training samples, then utilizes a saliency map to identify crucial regions in the training samples for backdoor injection.


\textbf{Adversarial attack}
in VFL involves an attacker following the VFL protocol during training but manipulating inputs at the inference phase to make the VFL model predict a specific targeted label.
Qiu et al. \cite{qiu2024hijack} propose HijackVFL, where the attacker uses a labeled dataset to train an auxiliary model. 
This model optimizes malicious feature embeddings that mislead the VFL model during collaborative inference. 
Gu et al. \cite{gu2023lrba} extend this idea with LR-BA, in which the attacker uses labeled training samples and applies the MC attack to train an additional model that generates malicious feature embeddings.
Pang et al. \cite{pang2023adi} design Adversarial Dominating Inputs (ADI) that dominate the joint inference process, directing it toward the attacker's desired outcome while reducing the contributions of honest participants. 
He et al. \cite{he2023backdoor} assume the attacker uses a large number of labeled samples (e.g., 500) from the target class to generate a trigger feature embedding. 
During inference, the attacker replaces their local embedding with this trigger feature embedding, causing the VFL model to misclassify any input as the target class.

\add{


}

    Although these methods demonstrate that an attacker can mislead the global VFL model to output a targeted label, their effectiveness relies on strong assumptions. 
    Specifically, these methods assume that the attacker has access to labeled samples in the training set. 
    However, in real-world VFL applications, this assumption is unrealistic since labels are critical private assets of the active party.
    Moreover, these methods overlook the fact that malicious feature embeddings can be easily detected if the active party deploys private anomaly detection mechanisms.
    In contrast, in this paper, we address a more practical scenario where the attacker has minimal adversarial knowledge and no access to training sample labels. 
    Additionally, we consider that the active party may employ anomaly detection techniques to identify adversarial embeddings from passive parties. 
    Even in this challenging and realistic setting, our framework effectively misleads the VFL model to predict test samples as targeted label.

\section{Conclusion}


In this paper, we propose VTarbel, a two-stage attack framework designed for targeted label attacks against detector-enhanced VFL inference systems, requiring minimal adversarial knowledge. 
The framework consists of two stages: in the preparation stage, the attacker selects a minimal number of highly expressive samples for benign VFL inference and uses the predicted labels from these samples to locally train an estimated detector model and a surrogate model. 
In the attack stage, the attacker employs gradient-based optimization techniques to generate malicious samples based on the two models trained in the preparation stage, and then transfers these optimized samples to the VFL inference system to launch the attack.
Comprehensive experiments demonstrate that VTarbel consistently outperforms other attack methods and remains resilient to representative defense strategies.

\bibliographystyle{ACM-Reference-Format}
\bibliography{reference}

@article{yang2019federated,
  title={Federated machine learning: Concept and applications},
  author={Yang, Qiang and Liu, Yang and Chen, Tianjian and Tong, Yongxin},
  journal={ACM Transactions on Intelligent Systems and Technology (TIST)},
  volume={10},
  number={2},
  pages={1--19},
  year={2019},
  publisher={ACM New York, NY, USA}
}

@misc{liu2022vertical,
    title={Vertical Federated Learning: Concepts, Advances and Challenges},
    author={Yang Liu and Yan Kang and Tianyuan Zou and Yanhong Pu and Yuanqin He and Xiaozhou Ye and Ye Ouyang and Ya-Qin Zhang and Qiang Yang},
    year={2022},
    eprint={2211.12814},
    archivePrefix={arXiv},
    primaryClass={cs.LG}
}

@article{li2023vertical,
  title={Vertical Federated Learning: Taxonomies, Threats, and Prospects},
  author={Li, Qun and Thapa, Chandra and Ong, Lawrence and Zheng, Yifeng and Ma, Hua and Camtepe, Seyit A and Fu, Anmin and Gao, Yansong},
  journal={arXiv preprint arXiv:2302.01550},
  year={2023}
}

@article{yang2023survey,
  title={A Survey on Vertical Federated Learning: From a Layered Perspective},
  author={Yang, Liu and Chai, Di and Zhang, Junxue and Jin, Yilun and Wang, Leye and Liu, Hao and Tian, Han and Xu, Qian and Chen, Kai},
  journal={arXiv preprint arXiv:2304.01829},
  year={2023}
}

@article{khan2022vertical,
  title={Vertical Federated Learning: A Structured Literature Review},
  author={Khan, Afsana and Thij, Marijn ten and Wilbik, Anna},
  journal={arXiv preprint arXiv:2212.00622},
  year={2022}
}

@article{wei2022vertical,
  title={Vertical federated learning: Challenges, methodologies and experiments},
  author={Wei, Kang and Li, Jun and Ma, Chuan and Ding, Ming and Wei, Sha and Wu, Fan and Chen, Guihai and Ranbaduge, Thilina},
  journal={arXiv preprint arXiv:2202.04309},
  year={2022}
}

@inproceedings{fu2022label,
  title={Label inference attacks against vertical federated learning},
  author={Fu, Chong and Zhang, Xuhong and Ji, Shouling and Chen, Jinyin and Wu, Jingzheng and Guo, Shanqing and Zhou, Jun and Liu, Alex X and Wang, Ting},
  booktitle={31st USENIX Security Symposium (USENIX Security 22)},
  pages={1397--1414},
  year={2022}
}

@inproceedings{li2022label,
    title={Label Leakage and Protection in Two-party Split Learning},
    author={Oscar Li and Jiankai Sun and Xin Yang and Weihao Gao and Hongyi Zhang and Junyuan Xie and Virginia Smith and Chong Wang},
    booktitle={International Conference on Learning Representations},
    year={2022},
    url={https://openreview.net/forum?id=cOtBRgsf2fO}
}

@article{sun2022label,
  title={Label leakage and protection from forward embedding in vertical federated learning},
  author={Sun, Jiankai and Yang, Xin and Yao, Yuanshun and Wang, Chong},
  journal={arXiv preprint arXiv:2203.01451},
  year={2022}
}

@ARTICLE{zou2022label,
  author={Zou, Tianyuan and Liu, Yang and Kang, Yan and Liu, Wenhan and He, Yuanqin and Yi, Zhihao and Yang, Qiang and Zhang, Ya-Qin},
  journal={IEEE Transactions on Big Data}, 
  title={Defending Batch-Level Label Inference and Replacement Attacks in Vertical Federated Learning}, 
  year={2022},
  volume={},
  number={},
  pages={1-12},
  doi={10.1109/TBDATA.2022.3192121}
}

@INPROCEEDINGS{tan2022residue,
  author={Tan, Juntao and Zhang, Lan and Liu, Yang and Li, Anran and Wu, Ye},
  booktitle={2022 8th International Conference on Big Data Computing and Communications (BigCom)}, 
  title={Residue-based Label Protection Mechanisms in Vertical Logistic Regression}, 
  year={2022},
  volume={},
  number={},
  pages={356-364},
  doi={10.1109/BigCom57025.2022.00051}
}

@article{kang2022framework,
  title={A framework for evaluating privacy-utility trade-off in vertical federated learning},
  author={Kang, Yan and Luo, Jiahuan and He, Yuanqin and Zhang, Xiaojin and Fan, Lixin and Yang, Qiang},
  journal={arXiv preprint arXiv:2209.03885},
  year={2022}
}

@inproceedings{fu2023focusing,
  title={Focusing on Pinocchio's Nose: A Gradients Scrutinizer to Thwart Split-Learning Hijacking Attacks Using Intrinsic Attributes.},
  author={Fu, Jiayun and Ma, Xiaojing and Zhu, Bin B and Hu, Pingyi and Zhao, Ruixin and Jia, Yaru and Xu, Peng and Jin, Hai and Zhang, Dongmei},
  booktitle={NDSS},
  year={2023}
}

@inproceedings{he2019model,
  title={Model inversion attacks against collaborative inference},
  author={He, Zecheng and Zhang, Tianwei and Lee, Ruby B},
  booktitle={Proceedings of the 35th Annual Computer Security Applications Conference},
  pages={148--162},
  year={2019}
}

@inproceedings{luo2021feature,
  title={Feature inference attack on model predictions in vertical federated learning},
  author={Luo, Xinjian and Wu, Yuncheng and Xiao, Xiaokui and Ooi, Beng Chin},
  booktitle={2021 IEEE 37th International Conference on Data Engineering (ICDE)},
  pages={181--192},
  year={2021},
  organization={IEEE}
}

@article{weng2020privacy,
  title={Privacy leakage of real-world vertical federated learning},
  author={Weng, Haiqin and Zhang, Juntao and Xue, Feng and Wei, Tao and Ji, Shouling and Zong, Zhiyuan},
  journal={arXiv preprint arXiv:2011.09290},
  year={2020},
  publisher={CoRR}
}

@article{jin2021cafe,
  title={Cafe: Catastrophic data leakage in vertical federated learning},
  author={Jin, Xiao and Chen, Pin-Yu and Hsu, Chia-Yi and Yu, Chia-Mu and Chen, Tianyi},
  journal={Advances in Neural Information Processing Systems},
  volume={34},
  pages={994--1006},
  year={2021}
}

@inproceedings {gao2023pcat,
  author = {Xinben Gao and Lan Zhang},
  title = {{PCAT}: Functionality and Data Stealing from Split Learning by {Pseudo-Client} Attack},
  booktitle = {32nd USENIX Security Symposium (USENIX Security 23)},
  year = {2023},
  isbn = {978-1-939133-37-3},
  address = {Anaheim, CA},
  pages = {5271--5288},
  url = {https://www.usenix.org/conference/usenixsecurity23/presentation/gao},
  publisher = {USENIX Association},
  month = aug
}

@inproceedings{pasquini2021unleashing,
  author = {Pasquini, Dario and Ateniese, Giuseppe and Bernaschi, Massimo},
  title = {Unleashing the Tiger: Inference Attacks on Split Learning},
  year = {2021},
  isbn = {9781450384544},
  publisher = {Association for Computing Machinery},
  address = {New York, NY, USA},
  url = {https://doi.org/10.1145/3460120.3485259},
  doi = {10.1145/3460120.3485259},
  booktitle = {Proceedings of the 2021 ACM SIGSAC Conference on Computer and Communications Security},
  pages = {2113–2129},
  numpages = {17},
  location = {Virtual Event, Republic of Korea},
  series = {CCS '21}
}

@article{liu2020backdoor,
  title={Backdoor attacks and defenses in feature-partitioned collaborative learning},
  author={Liu, Yang and Yi, Zhihao and Chen, Tianjian},
  journal={arXiv preprint arXiv:2007.03608},
  year={2020}
}

@INPROCEEDINGS {naseri2024badvfl,
  author = { Naseri, Mohammad and Han, Yufei and De Cristofaro, Emiliano },
  booktitle = { 2024 IEEE Symposium on Security and Privacy (SP) },
  title = {{ BadVFL: Backdoor Attacks in Vertical Federated Learning }},
  year = {2024},
  volume = {},
  ISSN = {},
  pages = {2013-2028},
  doi = {10.1109/SP54263.2024.00008},
  url = {https://doi.ieeecomputersociety.org/10.1109/SP54263.2024.00008},
  publisher = {IEEE Computer Society},
  address = {Los Alamitos, CA, USA},
  month =May
}

@inproceedings {bai2023villain,
    author = {Yijie Bai and Yanjiao Chen and Hanlei Zhang and Wenyuan Xu and Haiqin Weng and Dou Goodman},
    title = {{VILLAIN}: Backdoor Attacks Against Vertical Split Learning},
    booktitle = {32nd USENIX Security Symposium (USENIX Security 23)},
    year = {2023},
    isbn = {978-1-939133-37-3},
    address = {Anaheim, CA},
    pages = {2743--2760},
    url = {https://www.usenix.org/conference/usenixsecurity23/presentation/bai},
    publisher = {USENIX Association},
    month = aug
}

@article{gu2023lrba,
  title={LR-BA: Backdoor attack against vertical federated learning using local latent representations},
  author={Gu, Yuhao and Bai, Yuebin},
  journal={Computers \& Security},
  volume={129},
  pages={103193},
  year={2023},
  publisher={Elsevier}
}

@ARTICLE{qiu2024hijack,
  author={Qiu, Pengyu and Zhang, Xuhong and Ji, Shouling and Li, Changjiang and Pu, Yuwen and Yang, Xing and Wang, Ting},
  journal={IEEE Transactions on Dependable and Secure Computing}, 
  title={Hijack Vertical Federated Learning Models As One Party}, 
  year={2024},
  volume={},
  number={},
  pages={1-18},
  doi={10.1109/TDSC.2024.3358081}
}

@INPROCEEDINGS {pang2023adi,
    author = {Pang, Qi and Yuan, Yuanyuan and Wang, Shuai and Zheng, Wenting},
    booktitle = {2023 IEEE Symposium on Security and Privacy (SP)},
    title = {ADI: Adversarial Dominating Inputs in Vertical Federated Learning Systems},
    year = {2023},
    volume = {},
    issn = {},
    pages = {1875-1892},
    doi = {10.1109/SP46215.2023.00172},
    url = {https://doi.ieeecomputersociety.org/10.1109/SP46215.2023.00172},
    publisher = {IEEE Computer Society},
    address = {Los Alamitos, CA, USA},
    month = {may}
}

@article{he2023backdoor,
  title={Backdoor Attack Against Split Neural Network-Based Vertical Federated Learning},
  author={He, Ying and Shen, Zhili and Hua, Jingyu and Dong, Qixuan and Niu, Jiacheng and Tong, Wei and Huang, Xu and Li, Chen and Zhong, Sheng},
  journal={IEEE Transactions on Information Forensics and Security},
  year={2023},
  publisher={IEEE}
}

@article{liu2022copur,
  title={CoPur: certifiably robust collaborative inference via feature purification},
  author={Liu, Jing and Xie, Chulin and Koyejo, Sanmi and Li, Bo},
  journal={Advances in Neural Information Processing Systems},
  volume={35},
  pages={26645--26657},
  year={2022}
}

@inproceedings{chen2021homomorphic,
  title={When homomorphic encryption marries secret sharing: Secure large-scale sparse logistic regression and applications in risk control},
  author={Chen, Chaochao and Zhou, Jun and Wang, Li and Wu, Xibin and Fang, Wenjing and Tan, Jin and Wang, Lei and Liu, Alex X and Wang, Hao and Hong, Cheng},
  booktitle={Proceedings of the 27th ACM SIGKDD Conference on Knowledge Discovery \& Data Mining},
  pages={2652--2662},
  year={2021}
}

@article{zhu2019deep,
  title={Deep leakage from gradients},
  author={Zhu, Ligeng and Liu, Zhijian and Han, Song},
  journal={Advances in neural information processing systems},
  volume={32},
  year={2019}
}

@inproceedings{basu2002semi,
  title={Semi-supervised clustering by seeding},
  author={Basu, Sugato and Banerjee, Arindam and Mooney, Raymond},
  booktitle={In Proceedings of 19th International Conference on Machine Learning (ICML-2002},
  year={2002},
  organization={Citeseer}
}

@article{krishna1999genetic,
  title={Genetic K-means algorithm},
  author={Krishna, K and Murty, M Narasimha},
  journal={IEEE Transactions on Systems, Man, and Cybernetics, Part B (Cybernetics)},
  volume={29},
  number={3},
  pages={433--439},
  year={1999},
  publisher={IEEE}
}

@incollection{NEURIPS2019_9015,
title = {PyTorch: An Imperative Style, High-Performance Deep Learning Library},
author = {Paszke, Adam and Gross, Sam and Massa, Francisco and Lerer, Adam and Bradbury, James and Chanan, Gregory and Killeen, Trevor and Lin, Zeming and Gimelshein, Natalia and Antiga, Luca and others},
booktitle = {Advances in Neural Information Processing Systems 32},
pages = {8024--8035},
year = {2019},
publisher = {Curran Associates, Inc.},
url = {http://papers.neurips.cc/paper/9015-pytorch-an-imperative-style-high-performance-deep-learning-library.pdf}
}

@article{chen2017tutorial,
  title={A tutorial on kernel density estimation and recent advances},
  author={Chen, Yen-Chi},
  journal={Biostatistics \& Epidemiology},
  volume={1},
  number={1},
  pages={161--187},
  year={2017},
  publisher={Taylor \& Francis}
}

@article{jia2024model,
  title={Model Pruning-enabled Federated Split Learning for Resource-constrained Devices in Artificial Intelligence Empowered Edge Computing Environment},
  author={Jia, Yongzhe and Liu, Bowen and Zhang, Xuyun and Dai, Fei and Khan, Arif and Qi, Lianyong and Dou, Wanchun},
  journal={ACM Transactions on Sensor Networks},
  year={2024},
  publisher={ACM New York, NY}
}

@inproceedings{deepae,
  title={Data poisoning attacks against autoencoder-based anomaly detection models: A robustness analysis},
  author={Bovenzi, Giampaolo and Foggia, Alessio and Santella, Salvatore and Testa, Alessandro and Persico, Valerio and Pescap{\'e}, Antonio},
  booktitle={ICC 2022-IEEE International Conference on Communications},
  pages={5427--5432},
  year={2022},
  organization={IEEE}
}

@inproceedings{he2016resnet,
  title={Deep residual learning for image recognition},
  author={He, Kaiming and Zhang, Xiangyu and Ren, Shaoqing and Sun, Jian},
  booktitle={Proceedings of the IEEE conference on computer vision and pattern recognition},
  pages={770--778},
  year={2016}
}

@article{simonyan2014vgg,
  title={Very deep convolutional networks for large-scale image recognition},
  author={Simonyan, Karen and Zisserman, Andrew},
  journal={arXiv preprint arXiv:1409.1556},
  year={2014}
}

@article{mnist_dataset,
  title={MNIST handwritten digit database},
  author={LeCun, Yann and Cortes, Corinna and Burges, CJ},
  journal={ATT Labs [Online]. Available: http://yann.lecun.com/exdb/mnist},
  volume={2},
  year={2010}
}

@article{cifar10_dataset,
  title={The CIFAR-10 dataset},
  author={Krizhevsky, Alex and Nair, Vinod and Hinton, Geoffrey},
  journal={online: http://www.cs.toronto.edu/kriz/cifar.html},
  volume={55},
  number={5},
  year={2014}
}

@online{fmnist_dataset,
  author       = {Han Xiao and Kashif Rasul and Roland Vollgraf},
  title        = {Fashion-MNIST: a Novel Image Dataset for Benchmarking Machine Learning Algorithms},
  date         = {2017-08-28},
  year         = {2017},
  eprintclass  = {cs.LG},
  eprinttype   = {arXiv},
  eprint       = {cs.LG/1708.07747},
}

@article{cinic10_dataset,
  title={Cinic-10 is not imagenet or cifar-10},
  author={Darlow, Luke N and Crowley, Elliot J and Antoniou, Antreas and Storkey, Amos J},
  journal={arXiv preprint arXiv:1810.03505},
  year={2018}
}

@inproceedings{imdb_dataset,
  title={Learning word vectors for sentiment analysis},
  author={Maas, Andrew and Daly, Raymond E and Pham, Peter T and Huang, Dan and Ng, Andrew Y and Potts, Christopher},
  booktitle={Proceedings of the 49th annual meeting of the association for computational linguistics: Human language technologies},
  pages={142--150},
  year={2011}
}

@article{pgd,
  title={Adversarial machine learning in network intrusion detection systems},
  author={Alhajjar, Elie and Maxwell, Paul and Bastian, Nathaniel},
  journal={Expert Systems with Applications},
  volume={186},
  pages={115782},
  year={2021},
  publisher={Elsevier}
}

@inproceedings{wolf-etal-2020-transformers,
    title = "Transformers: State-of-the-Art Natural Language Processing",
    author = "Thomas Wolf and Lysandre Debut and Victor Sanh and Julien Chaumond and Clement Delangue and Anthony Moi and Pierric Cistac and Tim Rault and Rémi Louf and Morgan Funtowicz and Joe Davison and Sam Shleifer and Patrick von Platen and Clara Ma and Yacine Jernite and Julien Plu and Canwen Xu and Teven Le Scao and Sylvain Gugger and Mariama Drame and Quentin Lhoest and Alexander M. Rush",
    booktitle = "Proceedings of the 2020 Conference on Empirical Methods in Natural Language Processing: System Demonstrations",
    month = oct,
    year = "2020",
    address = "Online",
    publisher = "Association for Computational Linguistics",
    url = "https://www.aclweb.org/anthology/2020.emnlp-demos.6",
    pages = "38--45"
}

@manual{python-socket,
  title = {Python Socket Library},
  author = {Python Software Foundation},
  year={1991},
  note = {Available at \url{https://docs.python.org/3/library/socket.html}},
}

@book{korte2011combinatorial,
  title={Combinatorial optimization},
  author={Korte, Bernhard H and Vygen, Jens and Korte, B and Vygen, J},
  volume={1},
  year={2011},
  publisher={Springer}
}

@inproceedings{inkawhich2019feature,
  title={Feature space perturbations yield more transferable adversarial examples},
  author={Inkawhich, Nathan and Wen, Wei and Li, Hai Helen and Chen, Yiran},
  booktitle={Proceedings of the IEEE/CVF Conference on Computer Vision and Pattern Recognition},
  pages={7066--7074},
  year={2019}
}

@article{gu2023survey,
  title={A survey on transferability of adversarial examples across deep neural networks},
  author={Gu, Jindong and Jia, Xiaojun and de Jorge, Pau and Yu, Wenqain and Liu, Xinwei and Ma, Avery and Xun, Yuan and Hu, Anjun and Khakzar, Ashkan and Li, Zhijiang and others},
  journal={arXiv preprint arXiv:2310.17626},
  year={2023}
}

@article{sanh2019distilbert,
  title={DistilBERT, a distilled version of BERT: smaller, faster, cheaper and lighter},
  author={Sanh, Victor and Debut, Lysandre and Chaumond, Julien and Wolf, Thomas},
  journal={arXiv preprint arXiv:1910.01108},
  year={2019}
}

@inproceedings{trec_dataset,
    title = "Learning Question Classifiers",
    author = "Li, Xin  and
      Roth, Dan",
    booktitle = "{COLING} 2002: The 19th International Conference on Computational Linguistics",
    year = "2002",
    url = "https://www.aclweb.org/anthology/C02-1150",
}

@article{liu2024pna,
  title={Pna: Robust aggregation against poisoning attacks to federated learning for edge intelligence},
  author={Liu, Jingkai and Lyu, Xiaoting and Duan, Li and He, Yongzhong and Liu, Jiqiang and Ma, Hongliang and Wang, Bin and Su, Chunhua and Wang, Wei},
  journal={ACM Transactions on Sensor Networks},
  year={2024},
  publisher={ACM New York, NY}
}

@article{mmd_definition,
  title={A kernel two-sample test},
  author={Gretton, Arthur and Borgwardt, Karsten M and Rasch, Malte J and Sch{\"o}lkopf, Bernhard and Smola, Alexander},
  journal={The Journal of Machine Learning Research},
  volume={13},
  number={1},
  pages={723--773},
  year={2012},
  publisher={JMLR. org}
}

@article{engel2004kernel,
  title={The kernel recursive least-squares algorithm},
  author={Engel, Yaakov and Mannor, Shie and Meir, Ron},
  journal={IEEE Transactions on signal processing},
  volume={52},
  number={8},
  pages={2275--2285},
  year={2004},
  publisher={IEEE}
}

@article{li2022vertical,
  title={Vertical semi-federated learning for efficient online advertising},
  author={Li, Wenjie and Xia, Qiaolin and Cheng, Hao and Xue, Kouyin and Xia, Shu-Tao},
  journal={arXiv preprint arXiv:2209.15635},
  year={2022}
}

\end{document}